\documentclass[twocolumn]{aastex631}

\let\tablenum\relax
\usepackage{siunitx}
\usepackage{gensymb}
\usepackage{amsmath}
\usepackage{multirow}
\usepackage{graphicx}
\usepackage{fontenc}
\usepackage{float}
\usepackage{ulem}
\usepackage{placeins}

\begin{document}

\newcommand{\jithu}[1]{\textcolor{purple}{#1}}


\title{Relative Influence of Coronal Properties of CME in their Evolution and Geo-effectiveness}

\author[0009-0002-9114-5880]{Jithu J Athalathil}
\affiliation{Department of Astronomy Astrophysics and Space Engineering, Indian Institute of Technology Indore, Khandwa Road, Simrol, Indore 453552, India}

\author[0000-0003-4442-5994]{Asif M Mandayapuram}
\affiliation{Department of Astronomy Astrophysics and Space Engineering, Indian Institute of Technology Indore, Khandwa Road, Simrol, Indore 453552, India}
\affiliation{Space Applications Centre, Indian Space Research Organisation, Ahmedabad 380015, India}

\author[0000-0001-8265-6254]{Prateek Mayank}
\affiliation{University Corporation for Atmospheric Research, CPAESS, Boulder, CO 80303, USA}
\affiliation{Space Weather TREC, University of Colorado, Boulder, CO 80303, USA
}

\author[0000-0001-5424-0059]{Bhargav Vaidya}
\affiliation{Department of Astronomy Astrophysics and Space Engineering, Indian Institute of Technology Indore, Khandwa Road, Simrol, Indore 453552, India}

\begin{abstract}
Uncertainties in the coronal mass ejection (CME) properties near the Sun remain a major challenge for accurate space weather forecasting. Understanding the sensitivity of CME evolution to the initial parameters is therefore essential for improving the predictions. We investigate the influence of physical CME properties prescribed at 0.1 AU on their subsequent heliospheric propagation and geoeffectiveness at Earth, using the SWASTi framework. We perform a systematically sampled ensemble of 400 global magnetohydrodynamic simulations by varying the initial CME density, speed, magnetic flux and temperature. We quantify the relative influence of these parameters on CME evolution, plasma properties at L1, and the resulting geomagnetic response. The analysis reveals a clear hierarchy of influence: initial speed primarily controls the peak radial velocity and arrival time at L1, magnetic flux controls the magnetic-field strength, and density governs the peak density and magnetopause compression. Primary parameters account for only half the output variance, highlighting the collective influence of other parameters. Density exhibits the broadest influence across the analyzed outputs. We find a mean arrival-time sensitivities of 2.43~h per $100~\mathrm{km\,s^{-1}}$ variation in injection speed and 1.99~h per $10^{-18}~\mathrm{kg\,m^{-3}}$ variation in injection density. The CME volume exhibits a two-phase evolution, characterized by an initial gradual expansion followed by a power-law regime with a median expansion index of 2.38. These results establish parameter-specific priorities for CME forecasting, highlighting the importance of constraining CME speed for arrival-time prediction, density for duration and pressure-driven effects, and magnetic flux for geomagnetic response at Earth.

\end{abstract}


\section{Introduction}

Coronal mass ejections (CMEs) are the primary drivers of severe space weather and major geomagnetic storms \citep{Gosling1990}. When Earth-directed, these massive eruptions of solar plasma and magnetic field can disrupt satellite operations, interrupt communication systems, and pose significant risks to ground-based power grids. Consequently, accurately forecasting the arrival time and geoeffectiveness of CMEs is a paramount objective for the space weather community.

Global 3D magnetohydrodynamic (MHD) models are the current operational standard for simulating the propagation of CMEs through the inner heliosphere. This includes the Wang-Sheeley-Arge-ENLIL (WSA-ENLIL) cone model \citep{Odstrcil2003}, the Space-Weather-Modeling Framework with the Alfv\'en Wave Solar Model (SWMF/AWSoM) \citep{Toth2005, Toth2012, vanderHolst2014}, the Magnetohydrodynamic Algorithm outside a Sphere (MAS/CORHEL) \citep{Mikic1999, Riley2001, Linker2024}, the heliospheric Lyon-Fedder-Mobarry model and its successor Grid Agnostic MHD for Extended Research Applications (LFM-Helio/GAMERA-Helio) \citep{Merkin2016, Provornikova2024}, Space-weather-forecast-Usable System Anchored by Numerical Operations and Observations - CME (SUSANOO-CME) \citep{Shiota2016}, European heliospheric forecasting information asset (EUHFORIA) \citep{Pomoell2018}, Icarus \citep{Verbeke2022}, and the Space Weather Adaptive Simulation framework (SWASTi) \citep{Mayank2022, Mayank2023, Mayank2025}.

These models typically require user-defined boundary conditions, injecting the CME structure at an inner heliospheric boundary based on parameters derived from relevant observations \citep{Odstrcil2004}. However, the complex nature of CME initiation and the limitations of remote sensing observations inherently introduce significant uncertainties into these initial input parameters \citep{Mierla2010, Verbeke2023}. Understanding how these uncertainties propagate through the heliosphere and ultimately identifying which parameters matter most for accurate Earth arrival predictions remains a fundamental challenge.

Previous studies have extensively explored the sensitivity of MHD models to variations in initial parameter space. \cite{Taktakishvili2010} and \cite{Falkenberg2010} investigated the sensitivity of the ENLIL model to initial solar input parameters, demonstrating that typical observational errors in CME speed, angular width, and directional trajectory lead to substantial deviations in predicted arrival times at Earth orbit. Building upon this foundation, \cite{Riley2021} and \cite{Millas2020} systematically analyzed the sources and magnitudes of uncertainty in global MHD predictions. They confirmed that the accuracy of arrival time forecasts is fundamentally limited by the fidelity of the initial input conditions, highlighting the urgent need to understand the nonlinear relationship between coronal CME parameters and the resulting plasma properties at L1.

To mitigate these inherent observational uncertainties, the forecasting community has increasingly adopted probabilistic approaches. This method relies on perturbing initial CME parameters to generate a distribution of arrival times \citep{Barnard2020}. The effectiveness of this perturbation-based framework depends critically on identifying the input parameters that exert the strongest influence on CME propagation. Therefore, assessing the relative importance of the initial CME parameters is essential for developing an informed and physically motivated probabilistic forecasting strategy. The necessity of understanding parameter sensitivity is further amplified during complex solar events. For instance, \cite{Lee2015} demonstrated the utility of ensemble modeling for successive halo CMEs, while \cite{Koehn2022} highlighted how interacting CMEs can compound initial uncertainties, fundamentally altering the kinematic evolution of the ejecta and occasionally resulting in perfect storms.

Furthermore, accurately predicting the geoeffectiveness of a CME requires a detailed understanding of the CME’s evolving magnetic structure and its interaction with the ambient solar wind, which ultimately governs the strength of the geomagnetic response at Earth. As recently demonstrated by \cite{Wang2026}, the orientation of erupting magnetic flux ropes evolves significantly from the solar corona to 1 AU. Because the strength and duration of the southward magnetic field component ($B_z$) dictate the coupling of the CME with Earth's magnetosphere, forecasting geoeffectiveness is highly sensitive to the initial plasma and magnetic boundary assumptions.

While previous studies have established the importance of initial CME parameters in determining forecast uncertainties, sensitivity analyses employing full 3D MHD models have necessarily been restricted to relatively small ensembles, typically ranging from a handful of runs exploring a single parameter \citep{Taktakishvili2010, Falkenberg2010, Scolini2018} to a few tens of members per event in operational settings \citep{Mays2015}. Larger ensembles have so far been feasible only with reduced-physics models \citep{Dumbovic2018, Barnard2020}, which cannot self consistently capture the CME-solar wind interaction. Consequently, statistically robust quantitative assessments of the relative importance of individual parameters, as well as their interactions and combined effects, remain limited.  A larger and more comprehensive set of simulations is therefore needed to systematically explore the CME parameter space and obtain statistically meaningful estimates of parameter sensitivities. 

In this study, we address this limitation by performing a systematic sensitivity analysis of CME input parameters at the 0.1 AU boundary using a substantially larger ensemble of simulations. By comprehensively sampling the relevant parameter space, we quantify the individual and combined influence of the initial plasma and kinematic parameters on CME propagation, arrival time, and geoeffectiveness at L1. This approach enables us to establish a more robust hierarchy of parameter sensitivities and identify the initial CME properties that most strongly govern the resulting space weather response.

The paper is organized as follows. Section \ref{sec:Methodology} describes the methodology, including the model used in this study, the sampling strategy adopted for the simulations, and the calculation of the output parameters. Section \ref{sec:Ensemble_results} presents the results obtained from the numerical simulations. Section \ref{sec:Sensitivity_analysis} focuses on the sensitivity analysis used to quantify the influence of the CME input parameters on the simulated outputs. Section \ref{sec:Discussions} discusses the key findings and their implications for CME propagation and geoeffectiveness and section \ref{sec:Conclusions} summarizes the main conclusions of this study.

\section{Methodology}
\label{sec:Methodology}

SWASTi is a physics-based 3D heliospheric modeling framework developed to simulate the propagation of solar wind and CMEs from the inner heliosphere to 1 AU and beyond. The background solar wind is constructed using a combination of the Potential Field Source Surface (PFSS) model \citep{Altschuler1969,Ballegooijen2000, Stansby2020} and the WSA model \citep{Arge2003}, which provide the coronal solar wind plasma conditions at the inner boundary (0.1~AU). CMEs are modeled using the FRi3D model \citep{Isavnin2016}, which describes the 3D geometry and magnetic structure of a CME flux rope. The CME obtained from FRi3D is injected on the background solar wind solution at the inner boundary and subsequently propagated through the heliosphere. This enables SWASTi to capture the interaction and evolution of CME structures within the structured solar wind background and their propagation toward the near-Earth environment. A detailed description of the SWASTi framework, including its solar wind modeling, CME implementation, and numerical formulation, is provided in \cite{Mayank2022,Mayank2023,Mayank2025}.

Although the FRi3D model provides a physically motivated 3D representation of the CME flux rope, its parameterization involves simplifying assumptions. In particular, the density and temperature within the FRi3D CME are prescribed as spatially uniform, whereas the plasma velocity varies radially according to the local toroidal expansion of the flux-rope geometry.
The FRi3D parameterization includes both morphological parameters, which describe the geometry and orientation of the CME, and physical parameters, such as propagation speed and magnetic flux.
The CME model is constructed broadly by first defining shape of the CME using a set of morphological parameters (viz-a-viz half-width, half-height, flattening and pancaking) and then populating it with a twisted flux rope of magnetic fields defined by flux, twist, chirality and polarity.
In the present study, we focus specifically on the influence of these physical parameters on CME propagation and geoeffectiveness, while keeping the morphological parameters fixed.
Among the magnetic field parameters, we only change flux as the other parameters are observed not to have any significant effects on studied parameters.
We also note that current formulation of FRi3D used in this study is different from that of original in \cite{Isavnin2016} and new description can be found in \cite{maharana_updates_2023}. The technique used to sample these physical parameters is described in the following section.

We used the ADAPT (2024-10-02 12:00) synoptic magnetogram as the observational input for the background solar wind model. We implement the PFSS model and the modified WSA prescription as outlined in Section 2.1 of \cite{Mayank2022}. The WSA parameters are set to $\beta=1.25$, $V_{\min}=250\,\mathrm{km\,s^{-1}}$, and $V_{\max}=550\,\mathrm{km\,s^{-1}}$. The 3D MHD simulations span a spherical domain reaching from $0.1\text{ AU}$ to $2.1\text{ AU}$ radially, spanning $\pm 60^\circ$ in latitude and covering the full $2\pi$ radians azimuthally. The simulation grid consists of 150 radial cells, maintaining a constant angular spacing of $2^\circ$ across both latitudinal and azimuthal directions. Further details, including the MHD setup and boundary specifications, are provided in Section 2.2 of \cite{Mayank2022}. The CME is injected into the simulation domain at 2024-09-23 04:00 UT, centered at $3^\circ$ in both latitude and longitude. The flux-rope injection is configured with half-width = $45^\circ$, half-height = $30^\circ$, pancaking = $0.6$, tilt = $0^\circ$, flattening = $0.5$, and twist = $2$, with both chirality and polarity set to $1$.

For observational comparison, we use the Richardson-Cane near-Earth ICME catalog\footnote{Richardson-Cane near-Earth ICME catalog, available \url{https://doi.org/10.7910/DVN/C2MHTH}.} for the past 20-years period \citep{RichardsonCane2010, RichardsonCane2011}. The maximum solar wind speed and minimum value of the geomagnetic Dst index are taken directly from the catalog. The cataloged transit speed is used to estimate the transit time from $0.1$ AU to $1.0$ AU, assuming constant propagation speed.
The maximum ICME magnetic field strength and CME duration for the past 20 years is obtained from the HELIO4CAST ICMECAT catalog \citep{Moestl2026}.\footnote{HELIO4CAST ICMECAT, available at \url{https://helioforecast.space/icmecat}. The dataset version used in this study is available at \url{https://doi.org/10.6084/m9.figshare.6356420.v23}.}

\subsection{CME parameter sampling using the Morris method}

We employ the Morris one-at-a-time screening method \citep{Morris1991} to sample the input parameter space. This approach allows for an efficient exploration of the parameter space through both randomized and finite discretized sampling, helping to characterize the model's response to input variations (see Appendix~\ref{app:Morris} for more details).

The Morris method is applied to  CME injection parameters at 0.1 AU, with the number of parameters $k=4$; namely the density, temperature, speed, and magnetic flux. The specific ranges for these parameters are detailed in Table~\ref{tab:parameter_ranges}. The magnetic flux parameter spans from $10^{12}$ to $50\times 10^{12}$ Wb and is sampled on a logarithmic scale. All other parameters are sampled in a linear scale.

\begin{table}[h]
\centering
\caption{Parameter ranges adopted in this study.}
\label{tab:parameter_ranges}
\begin{tabular}{lc}
\hline
\textbf{Parameter} & \textbf{Range} \\
\hline
Density ($\times10^{-18}\,\mathrm{kg\,m^{-3}}$) & $0.5$--$10$ \\
Temperature (MK) & $0.3$--$1.5$ \\
Speed ($\mathrm{km\,s^{-1}}$) & $500$--$2000$ \\
Flux parameter ($\times 10^{12} $Wb) & $1-50$ \\
\hline
\end{tabular}
\end{table}



The sampling process utilizes 80 randomly generated parameter space trajectories in total. Each trajectory consists of $k+1=5$  model evaluations, yielding a comprehensive ensemble of 400 simulations.
To provide a baseline for our analysis, we designate a single ensemble member as the reference simulation, hereafter referred to as the ``reference run.'' This run corresponds to an initial CME density of $2\times 10^{-18} \,\mathrm{kg \,m}^{-3}$, a temperature of 0.8 MK, an initial speed of 2000 km s$^{-1}$, and a magnetic flux of $10\times 10^{12}$ Wb. While this representative simulation is highlighted throughout the text and figures to illustrate the behavior of the SWASTi model, it does not receive any weighted or special treatment during the statistical analysis.

\begin{figure*}[htbp]
    \centering
    \gridline{
        \fig{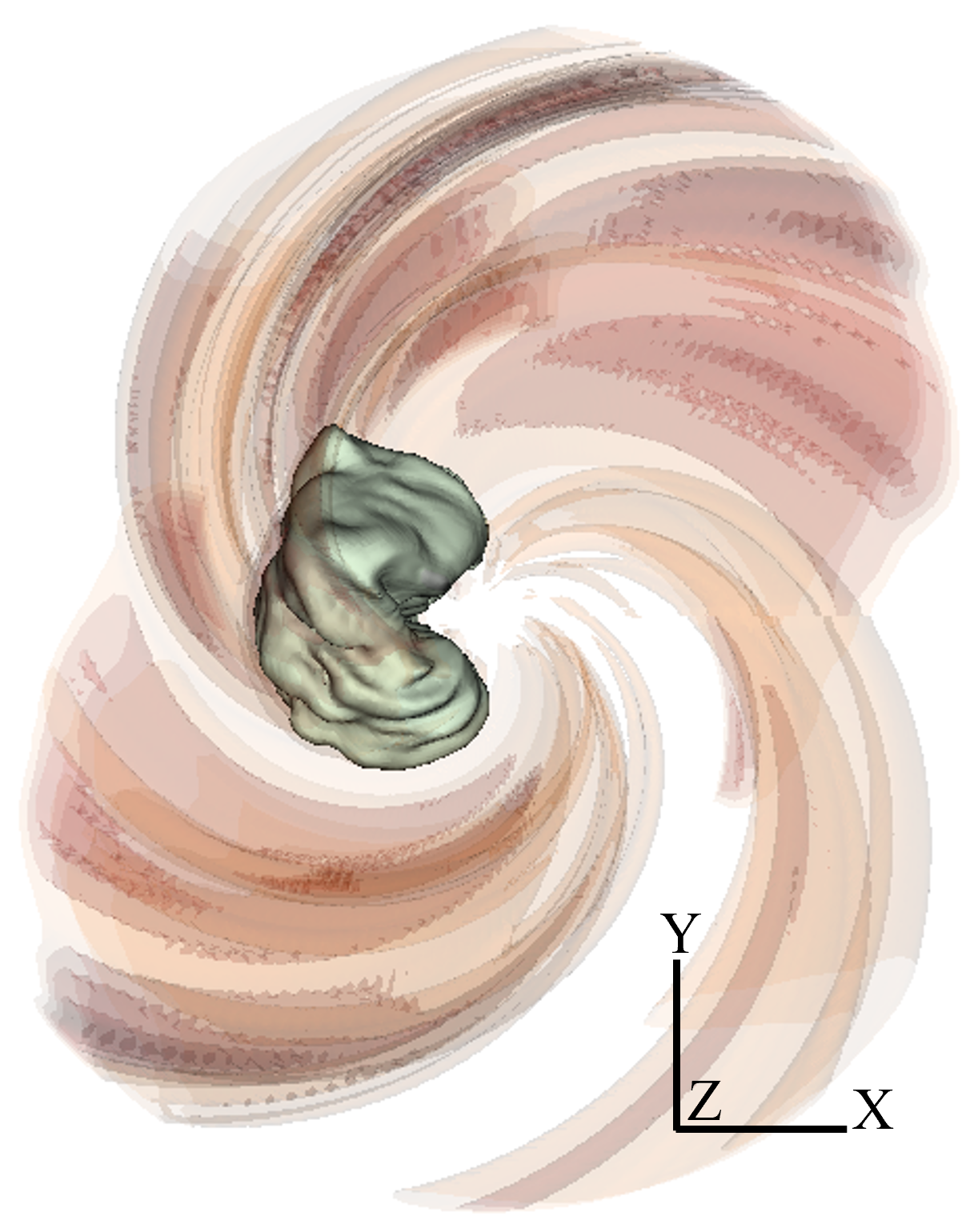}{0.28\textwidth}{}
        \fig{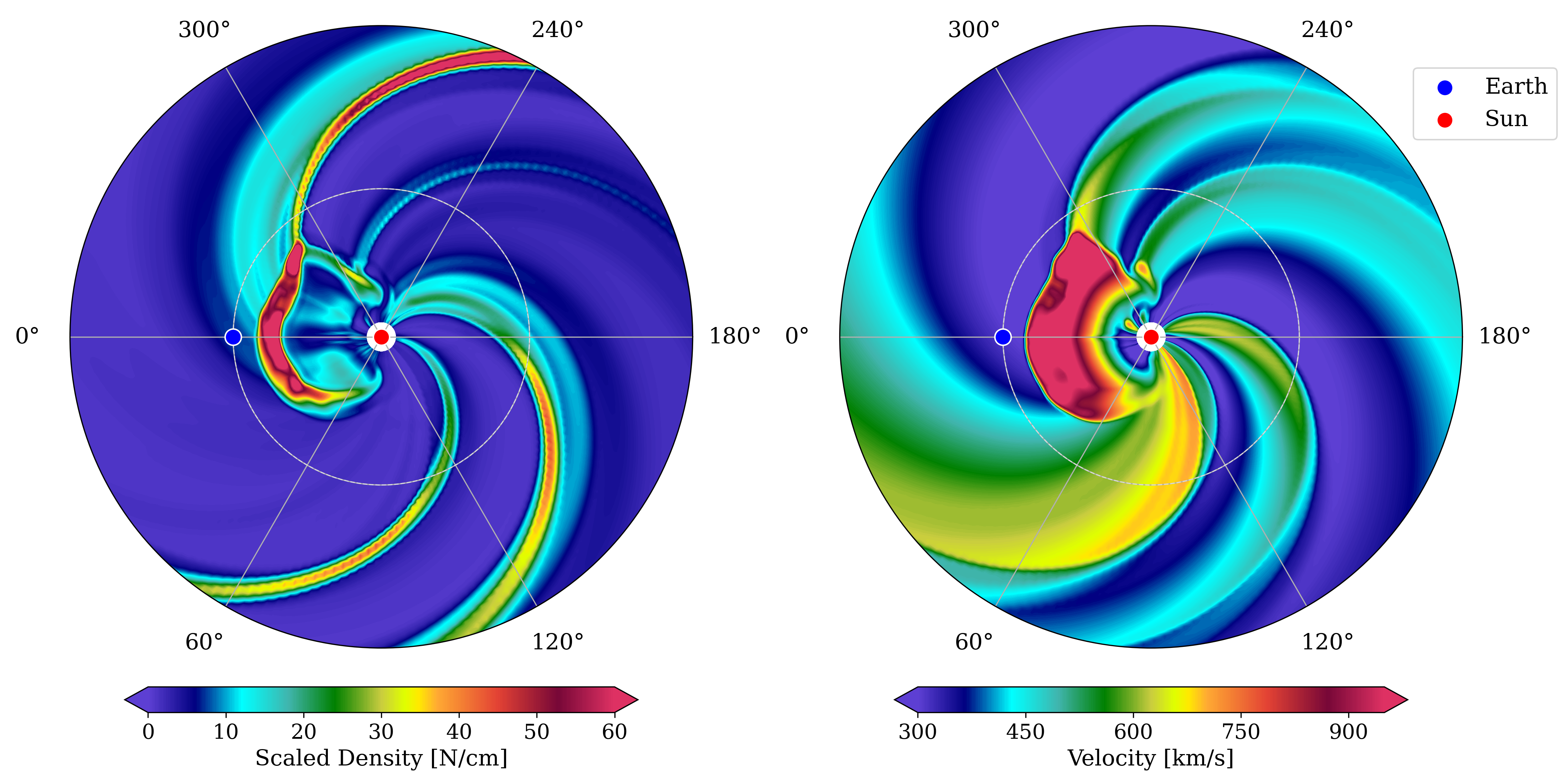}{0.67\textwidth}{}
    }
    \caption{Visualization of the CME propagating through the inner heliosphere. \textbf{Left}: 3D volume rendering of the CME tracer (green),
    overlaid on the scaled density of the ambient solar wind. \textbf{Middle}: scaled density in the ecliptic plane and \textbf{Right}: plasma velocity in the ecliptic plane. The CME appears as the high-density, high-velocity structure expanding away from the Sun, while the spiral bands of enhanced density and the alternating fast and slow streams correspond to the ambient solar wind and the SIR.}
    \label{fig:CME}
\end{figure*}

\begin{figure*}[htbp]
    \centering

    \gridline{
        \fig{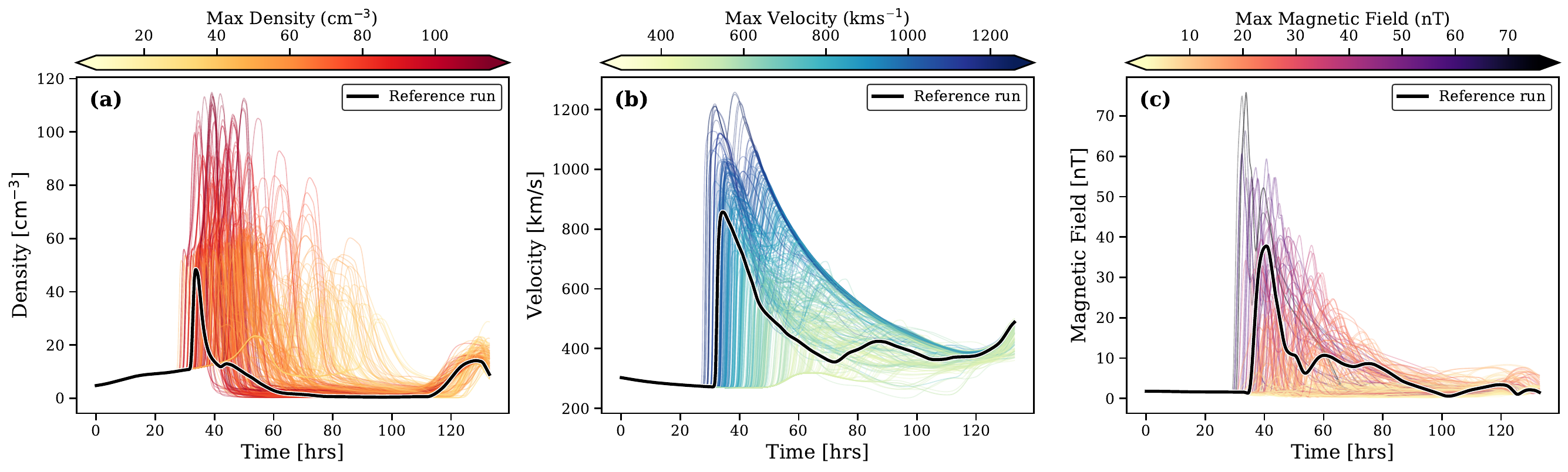}{0.9\textwidth}{}
    }

    \gridline{
        \fig{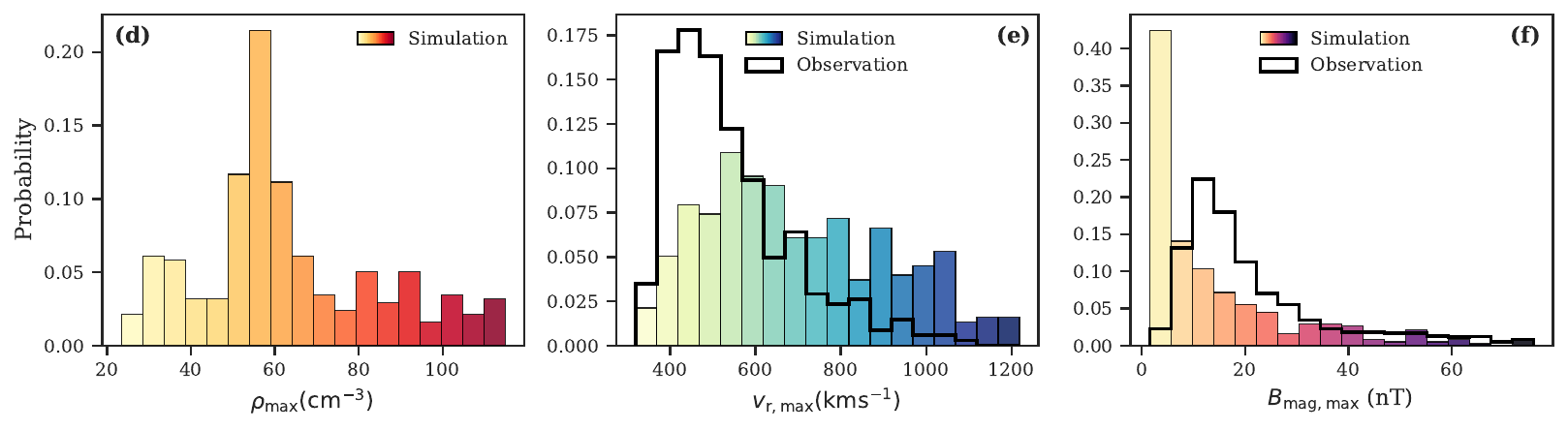}{0.9\textwidth}{}
    }

    \caption{
    Solar wind parameters at L1. \textbf{Top row :} The time series of (a) density, (b) velocity and (c) magnetic field at the L1. The lines represent individual simulations and the black curve corresponds to the reference simulation. For each simulation, the maximum value of each parameter is extracted and used as a scalar output metric for the subsequent statistical analysis. The individual simulations are color coded based on this peak value.
    \textbf{Bottom Row:} Distributions of (d) peak density ($\rho_{\max}$), (e) peak radial velocity ($V_{r,\max}$), and (f) peak magnetic field magnitude ($B_{\mathrm{mag},\max}$) obtained from the simulations. The distribution of maximum speeds from the Richardson--Cane catalog and maximum total magnetic field from ICMECAT is overplotted on the simulation results in panel(e and f) respectively for comparison.}
    \label{fig:L1_parameters}
\end{figure*}

\subsection{Magnetopause standoff distance}
\label{subsec:Magnetopause_distance}
The subsolar magnetopause position, which marks the boundary where the solar wind dynamic pressure is balanced by the Earth's magnetic field pressure, is determined using the empirical model of \cite{Shue1998}. The subsolar standoff distance $r_0$ (in units of Earth radii, $R_E$) is defined as

\begin{equation}
r_0 =
\left[
10.22 + 1.29 \tanh \left(0.184\left(B_z + 8.14\right)\right)
\right]
P_{\rm dyn}^{-1/6.6},
\end{equation}

where $P_{\rm dyn}$ is the solar wind dynamic pressure in nanopascals (nPa) and $B_z$ is the z-component of interplanetary magnetic field(IMF) in GSM coordinates, expressed in nanoteslas (nT). This model captures the nonlinear compression of the magnetosphere driven by varying solar wind conditions. To quantify the maximum magnetospheric compression during the CME impact, we utilized the minimum value of $r_0$ recorded over the simulation duration. It is important to note that the model is strictly valid only for dynamic pressures $P_{\mathrm{dyn}} > 0.5 \,\mathrm{nPa}$, which yields a maximum standoff distance of approximately 12 $R_E$. Values exceeding this distance are non-physical artifacts of the empirical formulation.

\subsection{Dst estimation}
\label{subsec:Dst_estimation}
The geoeffectiveness of the simulated CME is evaluated using the Disturbance Storm Time (Dst) index. The Dst index is estimated from the simulated solar wind parameters at 1 AU using the empirical model of \cite{OBrien2000} as described below.

The pressure-corrected Dst index, $Dst^{*}$, is given by

\begin{equation}
\frac{dDst^{*}}{dt} = Q(t) - \frac{Dst^{*}}{\tau},
\end{equation}

where $Q$ is the ring current injection term and $\tau$ is the ring current decay timescale. These quantities are calculated as

\begin{equation}
Q = 
\begin{cases}
-4.4 \left( VB_s - 0.5 \right), & VB_s > 0\\
0, & VB_s \leq 0
\end{cases}
\end{equation}

\begin{equation}
\tau = 2.4 \exp \left( \frac{9.74}{4.69 + VB_s} \right).
\end{equation}

Here, $VB_s$ is the solar-wind coupling parameter, expressed in $\mathrm{mV\,m^{-1}}$, and is defined as

\begin{equation}
VB_s =
\begin{cases}
\frac{|V B_z|}{1000}, & B_z < 0, \\ 0, & B_z \geq 0,
\end{cases}
\end{equation}

where $V$ is the solar wind speed in $\mathrm{km\,s^{-1}}$ and $B_z$ is the z-component of IMF in GSM coordinates, expressed in nT. The final Dst index is obtained after applying the magnetopause pressure correction,

\begin{equation}
Dst = Dst^{*} + 7.26 \sqrt{P_{\rm dyn}} - 11,
\end{equation}

where $P_{\rm dyn}$ is the solar wind dynamic pressure. The initial condition is set to $Dst = 0$ nT, corresponding to geomagnetically quiet conditions. The minimum of the resulting Dst profile is used to assess the geomagnetic impact of the simulated CME events.

To quantify the cumulative geomagnetic disturbance associated with each simulation run, the area under the Dst curve is calculated using the empirically determined Dst values. The integrated Dst area provides a combined measure of the magnitude and duration of Dst variations, thereby capturing the total geomagnetic disturbance represented by the Dst profile.

\subsection{CME Duration}
\label{subsec:CME_duration}
To identify CME plasma within the simulation domain, we employ a passive tracer variable $T_r$. The tracer is initialized with a value of unity inside the CME at the time of insertion and zero elsewhere in the computational domain. As a passive scalar, it is advected with the plasma flow without affecting the evolution of any physical quantities. The evolution of the tracer is governed by the conservative advection equation,

\begin{equation}
\frac{\partial T_r}{\partial t} + \nabla \cdot (T_r \mathbf{v}) = 0,
\end{equation}

where $\mathbf{v}$ is the plasma velocity. As the CME propagates and interacts with the ambient solar wind, numerical and physical mixing cause the tracer value to decrease from its initial value. Plasma with $T_r \geq 0.9$ is classified as CME magnetic cloud material. The CME duration corresponds to the total time during which CME cloud-associated plasma is detected at the Sun-Earth L1 point. The tracer threshold is chosen to be consistent with observations in which the CME duration is defined by the passage of the magnetic cloud. However the choice of tracer threshold can affect the estimated CME duration by including different portions of the CME in its evolution. Therefore the results and sensitivities related to CME duration should be interpreted with this in mind.

\subsection{Volume and energy density evolution}
\label{subsec:Volume_Energy_Evolution}

To quantify the spatial expansion and energetics of the CME, we isolated the ejecta from the ambient medium. Plasma with $T_r \geq 0.1$ is classified as CME material, whereas plasma with $T_r < 0.1$ is considered part of the ambient solar wind. The total volume of the CME is then calculated by summing the volumes of the selected grid cells at each time step. The temporal evolution of the CME volume is examined for each simulation, and the expansion index ($\gamma$) is determined by fitting a power-law relation, $V(t) \propto t^{\gamma}$, after excluding the initial part of the volume evolution.

Within the same set of cells, the temporal evolution of CME energetics is quantified by tracking the volume averaged magnetic ($E_{\rm mag}$), kinetic ($E_{\rm kin}$), and thermal ($E_{\rm th}$) energy densities at each time step. The evolution of each energy density is subsequently characterized by a power-law model of the form
\begin{equation}
E(t) \propto t^{-\beta},
\end{equation}
where $\beta$ represents the energy dissipation or decay index. For each simulation, $\beta_{\rm mag}$, $\beta_{\rm kin}$, and $\beta_{\rm th}$ are obtained from a least-square fit. To minimize the influence of numerical boundary effects, simulations are excluded from further analysis once the CME tracer reaches any of the domain boundaries. This helps to keep a check that the CME ejecta does not leave the computational domain.

\section{Ensemble results}
\label{sec:Ensemble_results}

In this section, we describe the output parameters derived from the CME ensemble simulations. These quantities characterize the CME evolution, its signatures at L1, and its geoeffectiveness, and are subsequently used in the sensitivity analyses.

Figure~\ref{fig:CME} illustrates a time stamp of the CME propagating through
the inner heliosphere using both 3D and 2D representations. The left panel
isolates the 3D topology of the transient using the CME tracer,
overlaid on the ambient scaled density. The two
accompanying polar plots provide an equatorial slice of the simulation domain, showing the scaled density (middle panel) and the solar wind velocity (right
panel) in the ecliptic plane. The positions of the Sun and Earth are marked for reference. The deformation in the CME front as it interacts with the SIR is clearly visible.

\begin{figure*}
    \centering
    \includegraphics[width=0.8\linewidth]{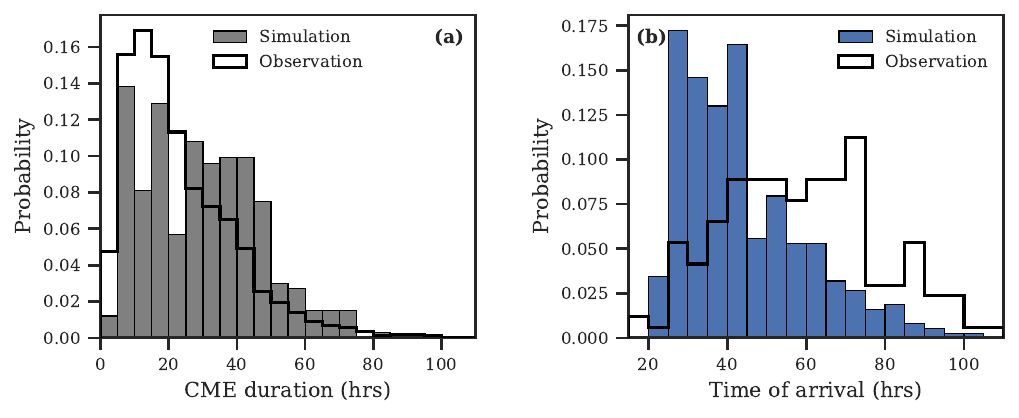}
    \caption{(a) Comparison of the CME duration distributions obtained from simulations and observations in the ICMECAT catalog. (b) Comparison of the CME transit-time distribution from simulations with the observational distribution derived from the Richardson–Cane catalog.}
    \label{fig:cme_duration}
\end{figure*}

\subsection{Plasma parameters at L1}

To quantify the L1 response of the CME ensemble, we use the time series of plasma density ($\rho$), radial velocity ($V_r$), and magnetic field magnitude ($B_{\mathrm{mag}}$) for each simulation. Figure~\ref{fig:L1_parameters} (top row) summarizes the temporal evolution of the L1 plasma parameters for the ensemble of CME simulations. Thin curves represent individual ensemble members, while the reference simulation is highlighted with a darker shade as a representative example. The same highlighting scheme is applied consistently across all parameters to aid the visual comparison of the CME signatures and their corresponding peak responses.

For every simulation, the maximum value of each parameter ($\rho_{\max}$, $V_{r,\max}$, and $B_{\mathrm{mag},\max}$) are extracted from the corresponding time series and used as a representative scalar measure of CME impact strength. The distributions of these derived metrics across the ensemble are shown in the bottom panels of Figure~\ref{fig:L1_parameters}. The ensemble distributions of $\rho_{\max}$, $V_{r,\max}$, and $B_{\mathrm{mag},\max}$ exhibit smooth and physically plausible ranges that are consistent with the diversity expected from CME propagation through the heliosphere. The corresponding L1 time series display the characteristic signatures of CME arrival, including density compression, velocity enhancement, and magnetic field amplification. A small number of simulations produce exceptionally large density or magnetic field peaks, appearing as tails in the distributions. These outliers correspond to scenarios with stronger compression and enhanced magnetic amplification. In the case of CMEs with high densities, we see reverse shocks building up giving a double peak structure.

The simulated distribution of maximum radial velocity is compared with the observed maximum ICME speeds from the Richardson--Cane catalog and magnetic field from ICMECAT (see Figure~\ref{fig:L1_parameters}(e and f)). The range of simulated maximum speeds agrees well with the observed range, indicating that the ensemble captures realistic velocity magnitudes. However, the overall distributions differ, as expected, because the input parameters were sampled randomly across the parameter space. Consequently, the relative frequencies in the simulated distribution do not necessarily represent the actual occurrence probabilities of CMEs. Observationally, lower-speed CMEs are more common than high-speed events, as the latter has greater energy requirements.

\begin{figure*}[htbp]
    \centering

    \gridline{
        \fig{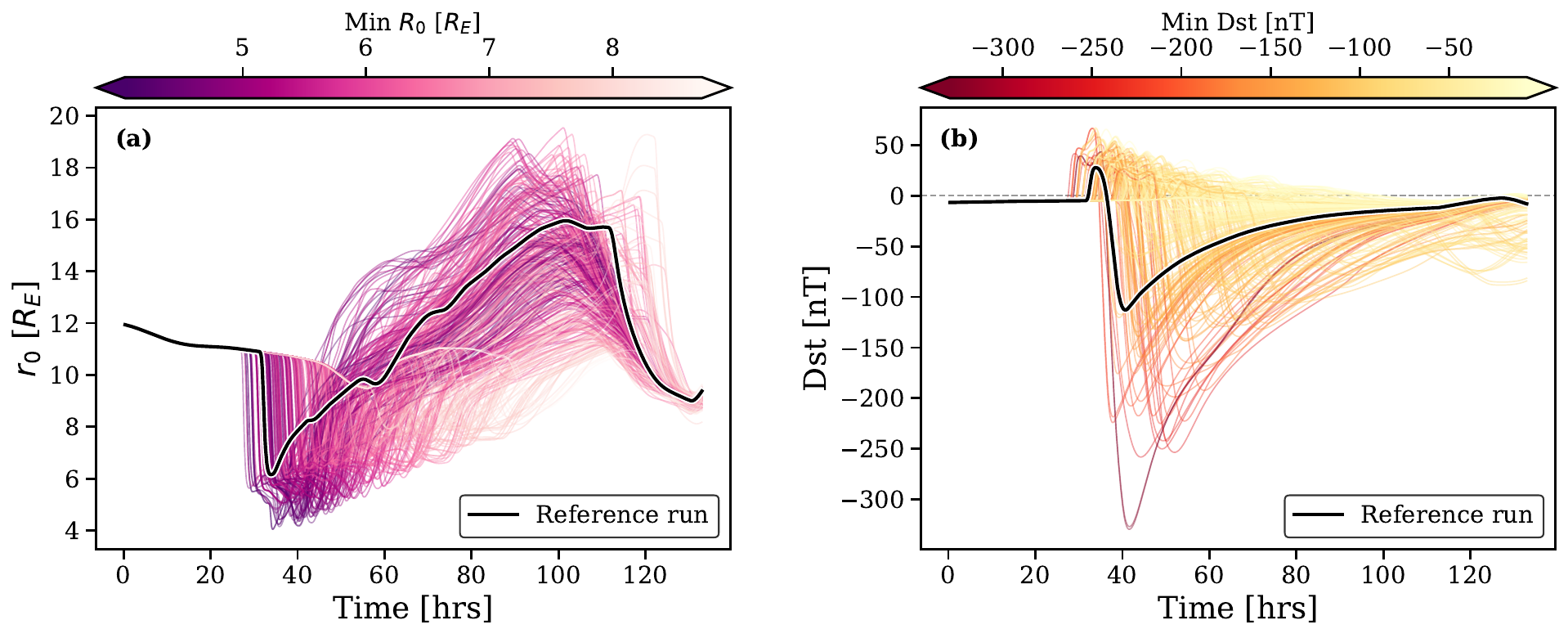}{0.9\textwidth}{}
    }

    \gridline{
        \fig{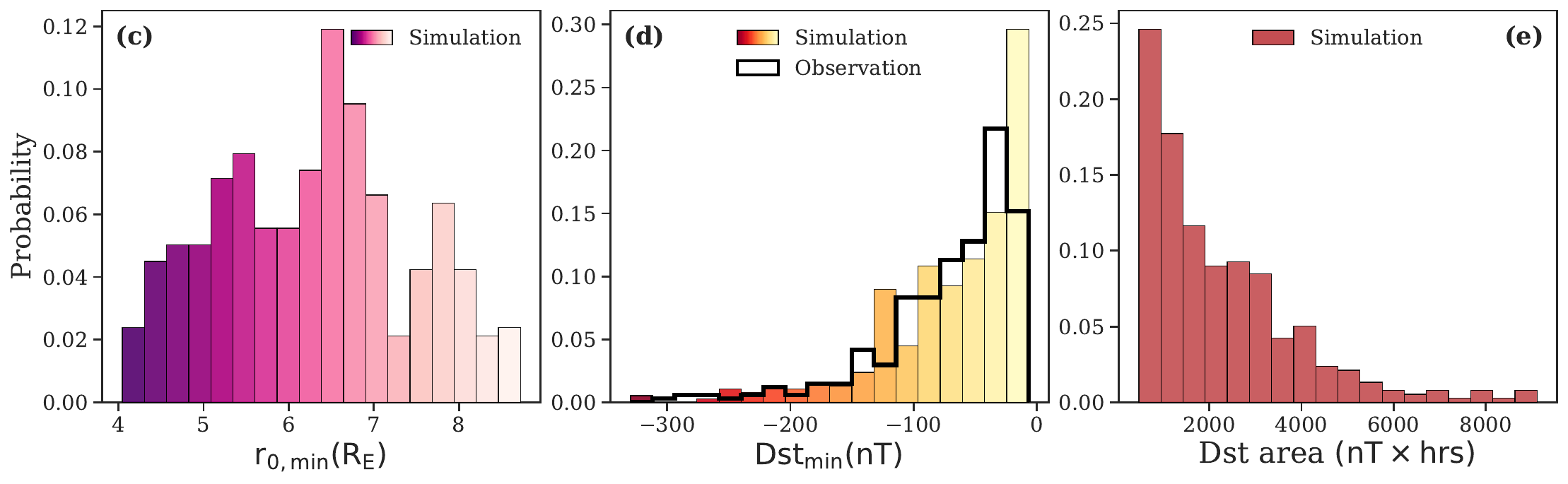}{0.9\textwidth}{}
    }
    
    \caption{
    Geoeffective parameters derived from the plasma parameters at L1.
    \textbf{Top row:} Temporal evolution of the modeled (a) subsolar magnetopause standoff distance ($r_0$) and (b) Dst index. For each simulation, the minimum magnetopause standoff distance ($r_{0,\min}$) and minimum Dst value ($\mathrm{Dst}_{\min}$) are extracted as scalar measures of the geomagnetic response. The ensemble members are color-coded according to these minimum values, while the reference simulation is shown in black.
    \textbf{Bottom row:} Distributions of (c) $r_{0,\min}$, (d) $\mathrm{Dst}_{\min}$, and (e) the integrated area under the Dst curve for all ensemble members. The observed $\mathrm{Dst}_{\min}$ values from the Richardson--Cane ICME catalog are overplotted in panel (d) for comparison.}
    \label{fig:Earth_parameters}
\end{figure*}

\subsection{CME duration and time of arrival at L1}

The CME duration at L1 is determined using the CME tracer, as described in Section~\ref{subsec:CME_duration}. The resulting distribution across the simulation ensemble is shown in Figure~\ref{fig:cme_duration}(a). The CME durations exhibit a pronounced positive skew, with most events persisting at L1 for approximately 10-40 hours. A broader tail extends toward longer durations, indicating that a smaller fraction of the simulated CMEs can remain detectable at L1 for nearly 60-80 hours. The observed CME duration distribution from the ICMECAT catalog is also overplotted in Figure~\ref{fig:cme_duration}(a), showing broad agreement with the simulated distribution in the overall duration range.

The difference between the CME injection time and the time at which the peak radial velocity is observed is used as a proxy for the CME arrival time at L1. Although shock arrival time can also serve as a proxy, we adopt the peak-radial-velocity time consistently across the ensemble to examine the statistical influence of the input parameters.
The resulting distribution of arrival times is shown in Figure~\ref{fig:cme_duration}(b). The ensemble exhibits a broad range of arrival times, demonstrating the strong sensitivity of CME transit time to the initial CME parameters. Most events reach L1 within approximately 25-50 hours, with the distribution peaking near 30 hours. A broader tail extends toward longer durations, indicating that a smaller fraction of the simulated CMEs travel slow. The observational arrival-time estimates derived from the Richardson--Cane catalog are overplotted in Figure~\ref{fig:cme_duration}(b), and their range agrees well with the simulated distribution. The higher representation of shorter arrival times is consistent with the shift in the simulated velocity distribution toward higher speeds, resulting in faster CME propagation to L1.

\subsection{Geo-effectiveness properties}

To assess the terrestrial response to the CME ensemble, we analyzed the temporal evolution of the subsolar magnetopause standoff distance ($r_0$) and the Dst index derived from each simulation as described in section~\ref{subsec:Magnetopause_distance} and ~\ref{subsec:Dst_estimation}. Figure~\ref{fig:Earth_parameters} (top row) illustrates the time series of these quantities for all ensemble members, with reference simulation highlighted.

The temporal evolution of both $r_0$ and Dst exhibits the expected behavior associated with CME-driven disturbances, including magnetospheric compression followed by recovery and the development of geomagnetic storm conditions. Most ensemble members produce moderate storm responses, while a limited number of simulations generate more extreme Dst depressions and larger integrated storm impacts. The smooth recovery phases and the absence of anomalous excursions in the time series provide confidence that the simulated magnetospheric response remains physically consistent throughout the ensemble.

For each simulation, the minimum magnetopause standoff distance ($r_{0,\min}$) and the minimum Dst value ($\mathrm{Dst}_{\min}$) are extracted from the corresponding time series and used as scalar measures of geomagnetic compression and storm intensity, respectively. In addition, the time-integrated area under the Dst curve is computed as a measure of the overall storm impact. These derived quantities constitute the primary Earth-impact metrics used in the subsequent statistical and sensitivity analyses. The distributions of $r_{0,\min}$, $\mathrm{Dst}_{\min}$, and the integrated Dst area across the ensemble are shown in the bottom row of Figure~\ref{fig:Earth_parameters}.

The observed minimum Dst from the Richardson--Cane catalog are overplotted in Figure~\ref{fig:Earth_parameters}(d), showing good agreement with the simulated distribution in both range and overall shape.
The logarithmic sampling of the input magnetic field strength provides relatively more samples at lower field strengths than uniform sampling in linear space, potentially contributing to the representation of moderate geomagnetic storms in the ensemble.

\begin{figure*}[htbp]
    \centering
    \includegraphics[width=0.75\linewidth,keepaspectratio]{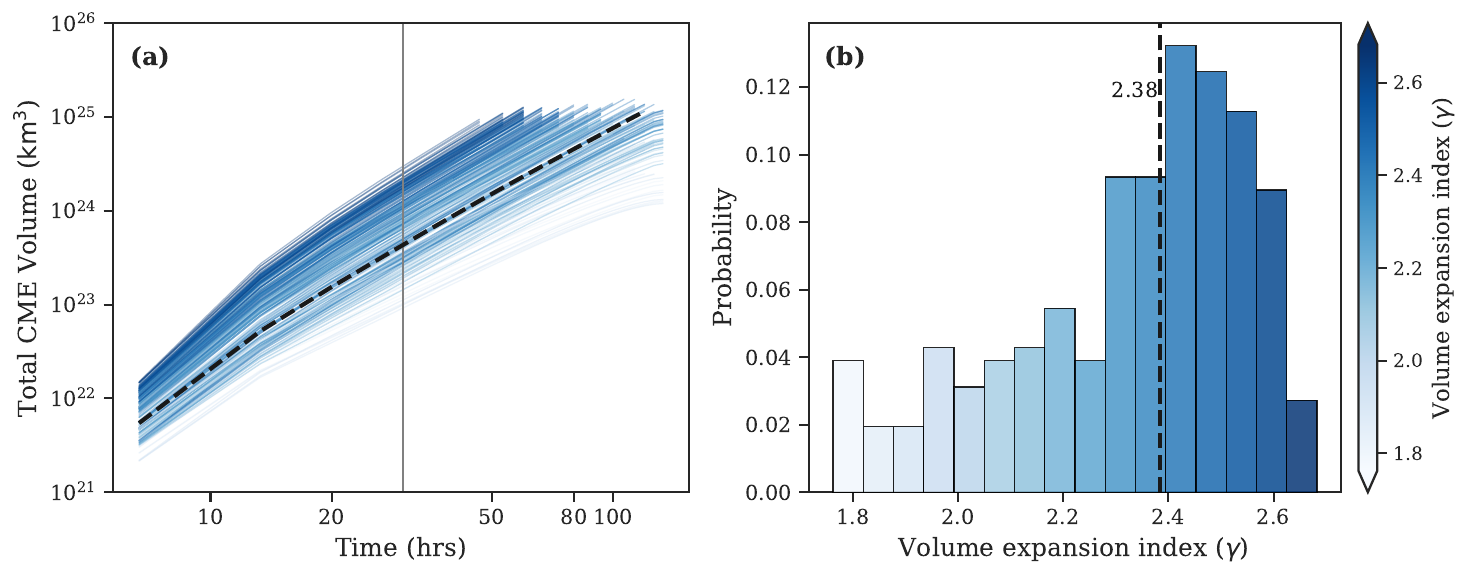}
    \caption{Volume evolution characteristics of the CME ensemble: (a) total CME volume as a function of time and (b) distribution of the expansion index ($\gamma$). The median $\gamma$ and the corresponding simulation are shown in black. Data for $t > 30\text{ hours}$ (marked by the gray vertical line) were used for the power law fit.}
    \label{fig:Volume_distribution}
\end{figure*}

\begin{figure*}[htbp]
    \centering

    \gridline{
        \fig{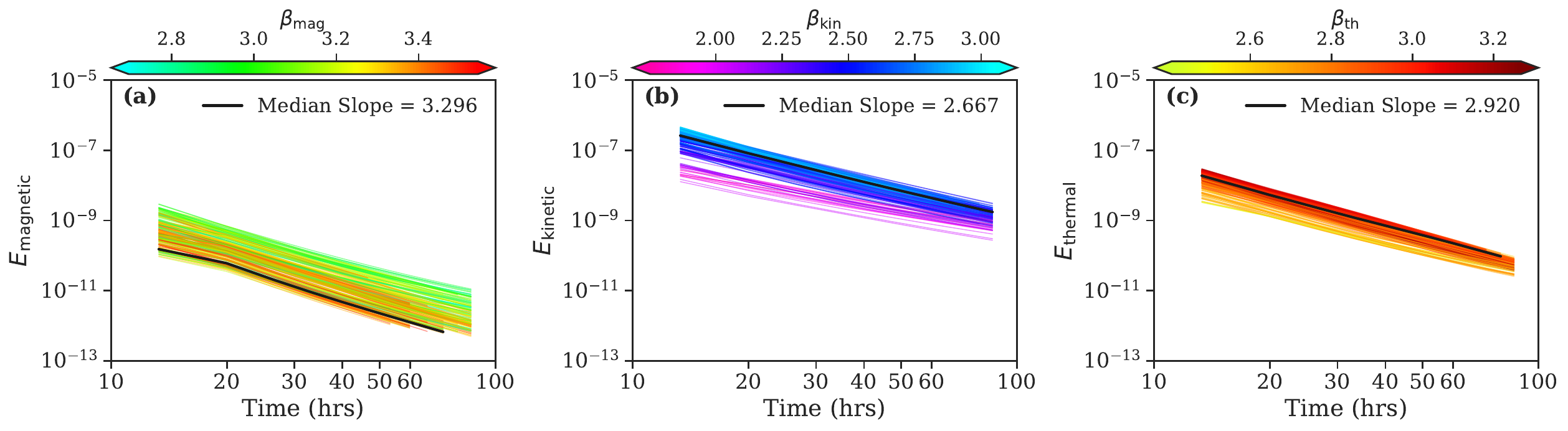}{0.98\textwidth}{}
    }

    \gridline{
        \fig{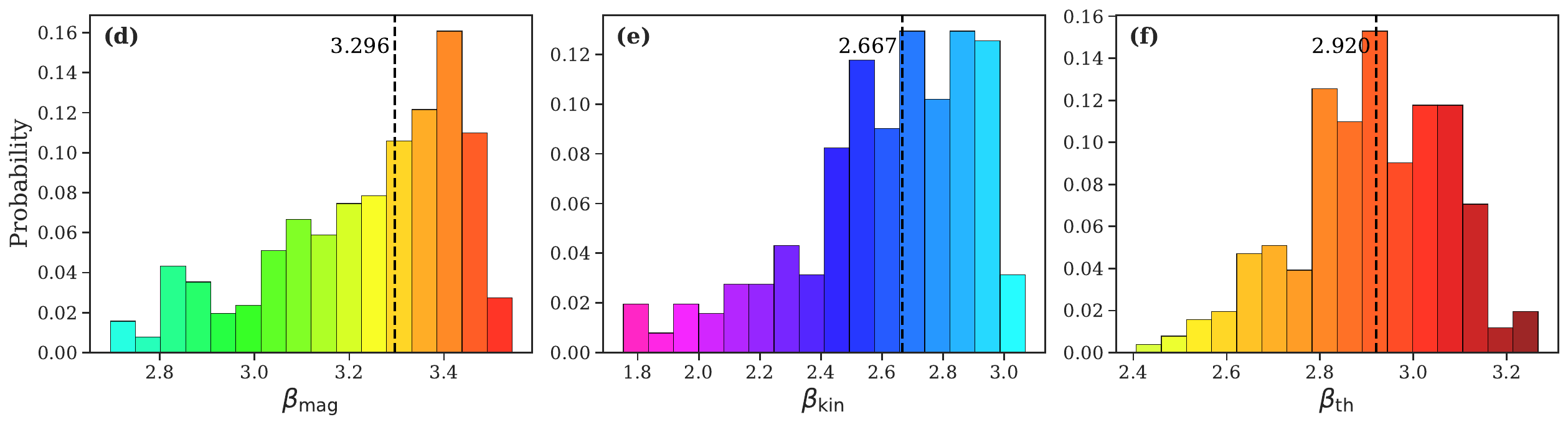}{0.95\textwidth}{}
    }

    \caption{
    Temporal energy density decay parameters.
    \textbf{Top row:} Temporal evolution of the (a) magnetic, (b) kinetic, and (c) thermal energy densities. The slope of each linear fit defines the respective decay-rate parameter, $\beta_{\mathrm{mag}}$, $\beta_{\mathrm{kin}}$, and $\beta_{\mathrm{th}}$. Individual ensemble members are color-coded according to the slope of the corresponding fit, while the simulation corresponding to the median slope are shown in black. 
    \textbf{Bottom row:} Distributions of the fitted decay-rate parameters (d) $\beta_{\mathrm{mag}}$, (e) $\beta_{\mathrm{kin}}$, and (f) $\beta_{\mathrm{th}}$ across the ensemble.
    }

    \label{fig:temporal_energy}
\end{figure*}

\subsection{Volume and energy density evolution}

Figure~\ref{fig:Volume_distribution} shows the volume evolution of the CME ensemble. The left panel presents the total CME volume as a function of time, revealing a two-phase expansion pattern. During the initial phase, the CME volume evolves gradually until approximately 15-30 hours, after which the ensemble exhibits a more systematic power-law increase. This two-phase evolution is consistent with the results reported by \cite{Mayank2023}. The volume evolution in the second phase is fitted with a power-law relation ($\rm{Volume} \propto \rm{t}^{\gamma}$) to determine the corresponding expansion index, $\gamma$. The right panel shows the distribution of $\gamma$, illustrating the statistical variability of the expansion indices across the CME ensemble. We find that the value of $\gamma$ ranges from 1.76 to 2.68, with a median value of $2.38_{-0.33}^{+0.16}$.

To quantify the temporal evolution of the CME energy budget, we analyzed the magnetic, kinetic, and thermal energy components for each ensemble member (see Section~\ref{subsec:Volume_Energy_Evolution}). Figure~\ref{fig:temporal_energy} (top row) shows the resulting log-log energy profiles for all ensemble members.

For each simulation, these energy components are modeled using power laws to obtain the decay-rate parameters $\beta_{\mathrm{mag}}$, $\beta_{\mathrm{kin}}$, and $\beta_{\mathrm{th}}$, respectively. The distributions of the derived decay-rate parameters across the ensemble are shown in the bottom row of Figure~\ref{fig:temporal_energy}. These results reveal a distinct hierarchy in the decay rates of CME energy densities during heliospheric propagation, with magnetic energy density dissipating the most rapidly ($\beta_{\mathrm{mag}} \approx 3.3$). Conversely, kinetic energy density decays at the slowest rate ($\beta_{\mathrm{kin}} \approx 2.7$), with thermal energy density falling in between ($\beta_{\mathrm{th}} \approx 2.9$).

\section{Sensitivity analysis}
\label{sec:Sensitivity_analysis}

In this section, we present a statistical sensitivity analysis of the ensemble outputs introduced in Section~\ref{sec:Ensemble_results} to identify which input parameters most strongly control the CME signatures at L1, and its geo-effectiveness. We use two methods to quantify sensitivity of output parameters to the inputs: i) Correlation which quantifies linear relationships ii) Sobol Variance which quantifies non linear contributions to overall outputs.

    

\subsection{Correlation}
\label{subsec:Correlation}

\begin{figure*}[!htbp]
    \centering

    \includegraphics[width=0.9\linewidth]{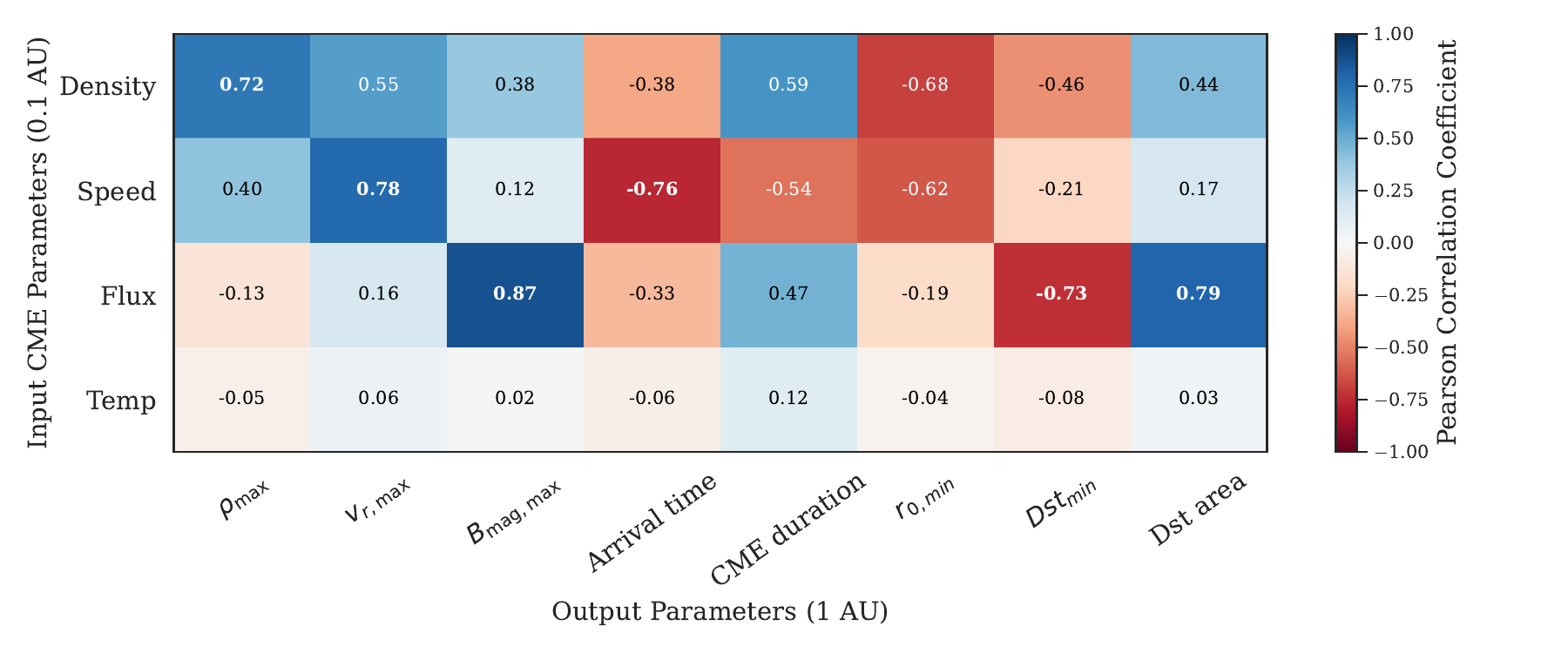}

    \vspace{0.5cm}

    \includegraphics[width=0.95\linewidth]{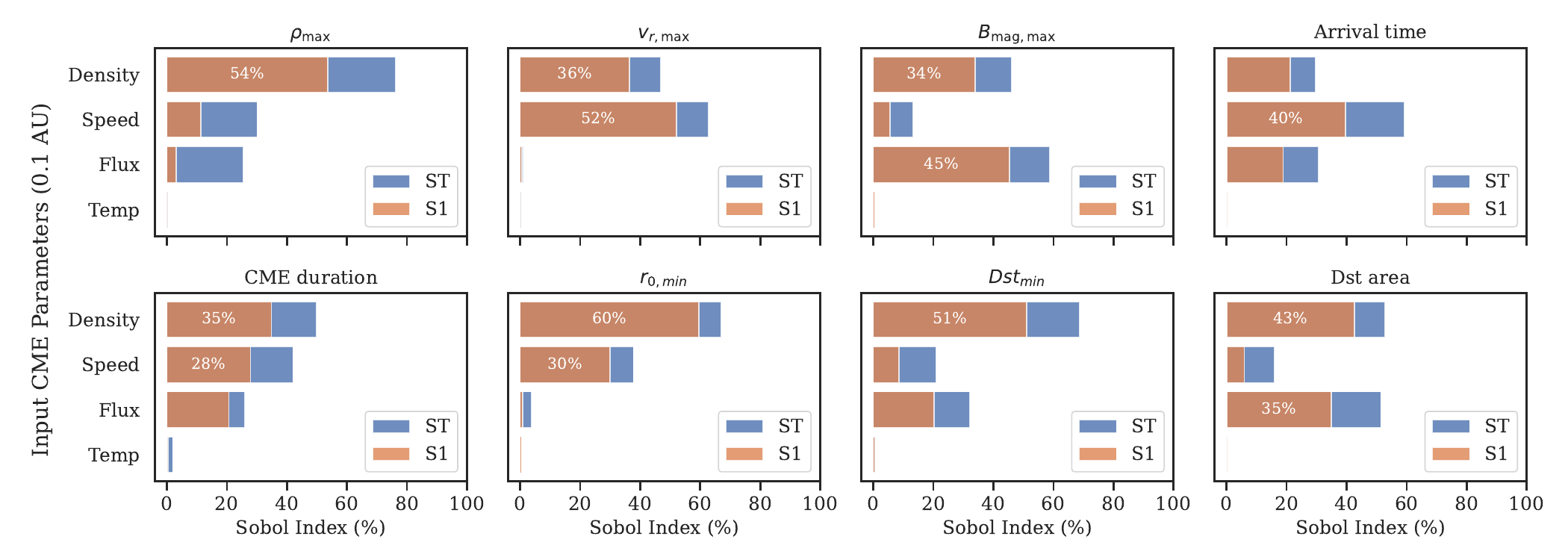}

    \caption{
    Correlation analysis and Sobol sensitivity analysis of the model.
    \textbf{Top row:} Correlation heatmap between input parameters and model outputs. The color scale represents the Pearson correlation coefficient.
    \textbf{Bottom row:} Sobol sensitivity indices for the output parameters. The first-order ($S_1$) and total-order ($S_T$) sensitivity indices are shown by orange and blue bars, respectively. The percentage contribution associated with the dominant source of variance is indicated on the corresponding bars. The values corresponding to all parameters are presented in Table~\ref{tab:sobol}.
    }
    \label{fig:correlation_sobol}
    \label{fig:correlation_heatmap}
    \label{fig:Sobol}
\end{figure*}

We quantified the linear dependence between the sampled input parameters and the model outputs using the Pearson correlation coefficient. The coefficient was computed for every input-output pair across the ensemble, and the resulting matrix is
shown in Figure~\ref{fig:correlation_heatmap}.

The strongest correlations reflect the expected primary physical dependencies inherent to the model. Density is strongly positively correlated with $\rho_{\max}$ ($r=0.72$) and moderately positively correlated with CME duration ($r=0.59$), indicating that denser eruptions tend to produce stronger density enhancements and longer-lasting CME signatures at L1. Speed shows a strong positive correlation with $v_{r,\max}$ ($r=0.78$) and a strong negative correlation with CME time of arrival ($r=-0.76$), indicating earlier arrival for faster cases. Speed shows a moderate negative correlation with CME duration ($r=-0.54$). Flux is strongly correlated with $B_{\mathrm{mag},\max}$ ($r=0.87$), consistent with its role in controlling magnetic field strength at L1. Flux also correlates negatively with the minimum Dst value ($r=-0.73$) and positively with the Dst area ($r=0.79$), implying that larger magnetic flux generally results in more intense geomagnetic storms.

Beyond these primary relationships, several secondary correlations are evident. Density exhibits moderate correlations with $v_{r,\max}$ ($r=0.55$) and $B_{\mathrm{mag},\max}$ ($r=0.38$), and a weak negative correlation with CME time of arrival ($r=-0.38$), suggesting a broader influence across multiple outputs. Temperature displays only weak correlations across all outputs, indicating limited linear influence within the sampled range.

\subsection{Sobol Variance decomposition}
Sobol variance decomposition is a variance-based global sensitivity analysis method that quantifies the fractional contribution of individual input parameters and their interactions to the overall variance of the model output. This approach has previously been employed to assess the importance of different parameters in the combined PFSS–WSA–Heliospheric Upwind eXtrapolation model \citep{Issan2023}. The method yields two primary metrics: the first-order sensitivity index ($S_1$), which measures the isolated main effect of a single parameter, and the total-order sensitivity index ($S_{T}$), which accounts for a parameter's main effect along with all its higher-order interactions with other parameters. A brief mathematical formulation of the Sobol decomposition is provided in Appendix~\ref{app:Sobol}. Unlike Pearson correlation, Sobol’s variance-based global sensitivity analysis captures nonlinear relationships and interactions across the input space.

Direct computation of Sobol sensitivity indices requires a large number of model evaluations far exceeding the number of simulations available in this study. We therefore constructed surrogate models using Random Forest regression, which is well suited to capturing nonlinear relationships between the input parameters and model outputs (see appendix \ref{app:RF} for more details). Separate Random Forest models are trained for each output quantity using the available simulation dataset and subsequently used to evaluate the model response over a large Sobol-sampled parameter space. The surrogate models reproduced most of the output parameters with high accuracy (cross-validated $R^2 >0.9$).

Figure~\ref{fig:Sobol} presents the Sobol sensitivity analysis for all output quantities. The values corresponding to all parameters are summarized in Table~\ref{tab:sobol}. The results show that each output most strongly influenced by the corresponding physical input parameter, with density providing the largest contribution to $\rho_{\max}$ and CME duration, speed having the strongest influence on $v_{r,\max}$ and CME time of arrival, and magnetic flux having the strongest influence on $B_{\mathrm{mag},\max}$. However, several outputs show appreciable contributions from multiple inputs. In particular, CME duration is influenced by density, speed, and magnetic flux, while the geoeffectiveness metrics depend on different combinations of density, speed, and magnetic flux.


\section{Discussion}
\label{sec:Discussions}

\subsection{Relative Influence of CME Properties}

The correlation and Sobol analyses reveal a clear hierarchy in the influence of CME injection parameters on heliospheric evolution and geoeffectiveness. The following subsections discuss the physical implications of these relationships for CME propagation and geoeffectiveness.

\subsubsection{Plasma parameters at L1}
The maximum density at 1 AU, $\rho_{\max}$, is most strongly influenced by the initial CME density (S1 = 54\%). CME speed provides a weaker secondary contribution, as faster CMEs additionally enhance the peak density through stronger compression during propagation. The maximum radial velocity, $V_{r,\max}$, is most strongly influenced by the initial CME speed (S1 = 52\%). The dominant influence of speed reflects the fact that differences in initial kinematics are largely preserved during propagation to 1 AU, rather than being erased by interaction with the background solar wind. The secondary dependence on density follows from the same drag-based mechanism. Denser CMEs carry greater inertia and are therefore decelerated less efficiently, allowing them to retain a larger fraction of their initial velocity by 1 AU. The maximum magnetic field magnitude at 1 AU, $B_{\mathrm{mag},\max}$, is most strongly influenced by the initial magnetic flux (S1 = 45\%). The injection density acts as a substantial secondary contributor. Thus, the magnetic field strength observed at L1 is determined principally by the magnetic content injected at 0.1 AU, while its subsequent evolution is also associated with the plasma density of the ejecta.

\subsubsection{Time of Arrival and CME Duration}
The initial CME speed has the strongest influence on time of arrival ($r=-0.76$, S1 = 40\%). Initial density and flux shows a negative correlation with time of arrival, indicating a secondary influence on CME propagation. These results establish speed as the leading contributor for arrival-time prediction within the present ensemble, while density and magnetic flux modifies the subsequent propagation through its influence on the evolution of the ejecta. In contrast, CME duration is most sensitive to the initial density ($S_1 = 35\%$), followed by initial speed ($S_1 = 28\%$) and magnetic flux ($S_1 = 21\%$).  Within the adopted tracer criterion, denser CMEs retain a larger radial extent of identifiable CME material, resulting in a longer passage duration at L1. Also a higher propagation speed may shorten the passage duration by allowing the CME to traverse L1 more rapidly.

\subsubsection{Impact on Earth}
The minimum magnetopause standoff distance, $r_{0,\min}$, is most strongly influenced by the initial CME density (S1 = 60\%) and speed (S1 = 30\%). CMEs that retain higher density and speed therefore produce stronger magnetopause compression and smaller values of $r_{0,\min}$. The $Dst_{\min}$, is most strongly correlated with the initial magnetic flux ($r=-0.73$). This relationship is consistent with the strong dependence of $B_{\mathrm{mag},\max}$ on magnetic flux and with the explicit dependence of the adopted Dst formulation on the southward IMF. Density also contributes substantially to $Dst_{\min}$ (S1 = 51\%). Thus, within the fixed magnetic configuration used in the ensemble, magnetic flux primarily determines the strength of the magnetic driver, while density influences the geomagnetic response through the plasma and pressure conditions delivered at L1. The integrated Dst response is also sensitive to both magnetic flux (r = 0.79, S1 = 35\%) and density (r =0.44, S1 = 43\%).  Magnetic flux governs the strength of the magnetic disturbance, whereas the density dependence of CME duration indicates how the disturbance can persist over a longer interval.


Temperature exhibits comparatively weak sensitivity across the investigated parameter range. It is also interesting to note that the outputs are not uniformly controlled by a single parameter. The ST indices indicate appreciable contributions from multiple inputs for several outputs, particularly density and speed for peak radial velocity and magnetopause compression, and density and magnetic flux for the Dst response. Consequently, the response of the CME and its geoeffectiveness reflect the combined influence of multiple injection parameters rather than the effect of a single parameter in isolation.



\subsection{Phases of CME Evolution}

In addition to the plasma and geomagnetic signatures at L1, the ensemble provides a statistical description of CME expansion and energy-density evolution throughout the inner heliosphere. The CME volume exhibits two distinct phases: an initial period of gradual evolution during the first approximately 15-30 hr, followed by a systematic power-law expansion. For (t$>$30) hr, the fitted expansion index ranges from 1.76 to 2.68, with a median value of $\gamma=2.38^{+0.16}_{-0.33}$. The occurrence of this behavior across the ensemble indicates that the two-phase evolution previously reported by \cite{Mayank2023} is retained over a broad range of initial CME density, temperature, speed, and magnetic flux. The difference between the present expansion index and that reported by \cite{Mayank2023} may arise from differences in the injected CME properties and their interaction with different background solar wind.

The expansion is accompanied by a systematic decrease in the volume-averaged magnetic, kinetic, and thermal energy densities. The corresponding median decay indices are $\beta_{\mathrm{mag}}=3.30$, $\beta_{\mathrm{kin}}=2.67$, and $\beta_{\mathrm{th}}=2.92$. It establishes that the magnetic energy density decreases most rapidly and the kinetic energy density most slowly. This ordering is consistent with the measured volumetric expansion. Under an approximately self-similar expansion, $V\propto R^{3}$, and magnetic flux conservation gives $B\propto R^{-2}$. The magnetic energy density would therefore scale as $B^{2}\propto V^{-4/3}$, which, for the median $\gamma=2.38$, corresponds to a temporal decay index of $4\gamma/3\approx3.17$. This value is close to the measured median $\beta_{\mathrm{mag}}=3.30$, indicating that the rapid decrease in magnetic energy density is largely consistent with magnetic field dilution during CME expansion.

For constant CME mass and slowly varying bulk velocity, the kinetic energy density would approximately follow the density dilution and scale as $V^{-1}$, corresponding to a decay index near the volume expansion index. The measured value $\beta_{\mathrm{kin}}=2.67$ is moderately larger than the median $\gamma=2.38$, consistent with expansion being accompanied by changes in the CME bulk velocity during its interaction with the ambient solar wind. The thermal energy density decay lies between the magnetic and kinetic components, indicating that its evolution is not governed by volume expansion alone. The fitted indices quantify the combined effects of expansion and heliospheric interaction. However, the present analysis does not separately determine the contributions from adiabatic evolution, compression, mixing, and numerical dissipation.

\subsection{Arrival-Time Sensitivity}
\label{sec:Uncertainity}

\begin{figure*}
    \centering
    \includegraphics[width=0.8\linewidth]{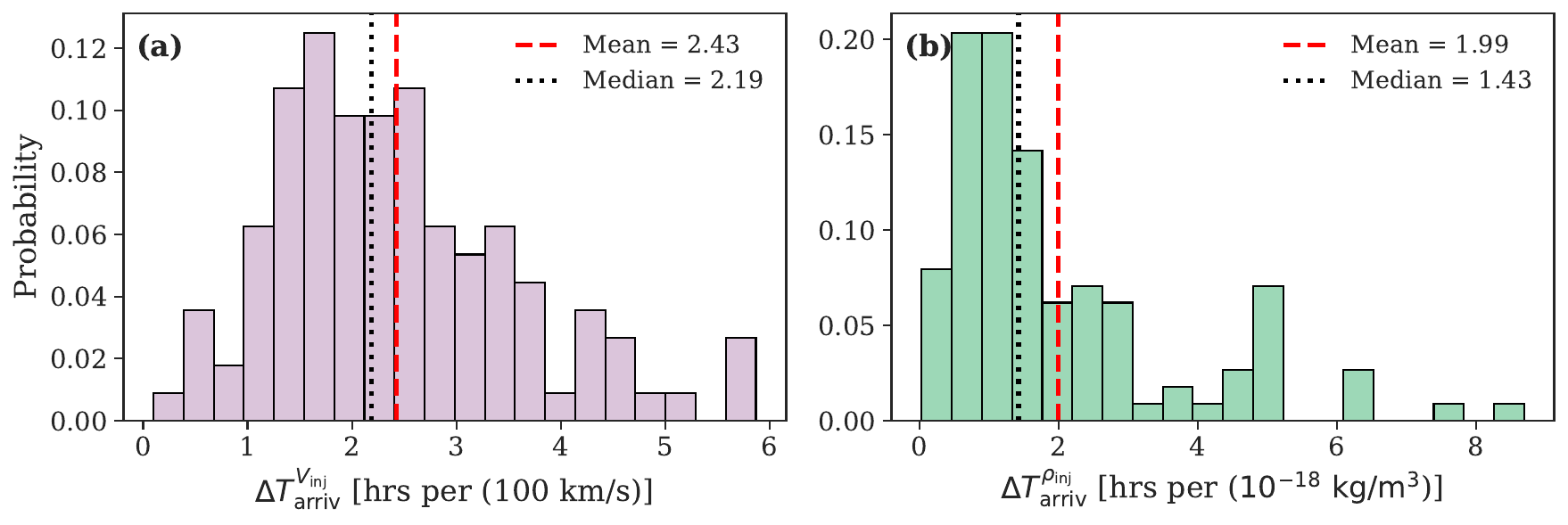}
    \caption{Probability distributions of arrival-time sensitivity to (a) CME injection speed, and (b) CME injection density. The red dashed and black dotted lines indicate the mean and median sensitivities, respectively.}
    \label{fig:Uncertainity}
\end{figure*}

Arrival-time prediction is a key objective of CME forecasting and, therefore, an important output of the simulation ensemble. To quantify the influence of CME injection speed and density on CME arrival time, we identify pairs of simulations in which only the corresponding input parameter changes, while all other parameters remain fixed. For each pair of simulations, the absolute change in arrival time, $|\Delta T_{\mathrm{arriv}}|$ (in  hrs), is divided by the corresponding change in the injection parameter to obtain a finite-difference sensitivity.
For injection speed, the sensitivity is normalized to a reference
increment of $100\,\mathrm{km\,s^{-1}}$ and is defined as

\begin{equation}
    \Delta T_{\mathrm{arriv}}^{V_{\mathrm{inj}}}
    =
    \left|
    \frac{\Delta T_{\mathrm{arriv}}}
    {\Delta v_{\mathrm{inj}}/(100\,\mathrm{km\,s^{-1}})}
    \right|.
\end{equation}
whereas the density sensitivity is normalized to a reference increment of $10^{-18}\,\mathrm{kg\,m^{-3}}$:

\begin{equation}
    \Delta T_{\mathrm{arriv}}^{\rho_{\mathrm{inj}}}
    =
    \left|
    \frac{
        \Delta T_{\mathrm{arriv}}
    }{
        \Delta \rho_{\mathrm{inj}} /
        (10^{-18}\,\mathrm{kg\,m^{-3}})
    }
    \right|.
\end{equation}
Here, $\Delta V_{\mathrm{inj}}$ (in $\mathrm{km\,s^{-1}}$) and $\Delta \rho_{\mathrm{inj}}$ (in $\mathrm{kg\,m^{-3}}$) represent the differences in injection speed and density between the paired simulations, respectively. The resulting sensitivities quantify the magnitude of the arrival-time response to variations in each input parameter. The distributions of these sensitivities across all eligible pairs are presented in Figure~\ref{fig:Uncertainity}.

The mean and median arrival-time sensitivities are 2.43 and 2.19 hours per $100\,\mathrm{km\,s^{-1}}$ for injection speed, respectively, compared with 1.99 and 1.43 hours per $10^{-18}\,\mathrm{kg\,m^{-3}}$ for injection density.
The calculated sensitivities provide a quantitative measure of how variations in the initial CME parameters can contribute to arrival-time uncertainty. They highlight the importance of accurately constraining both injection speed and density for reliable CME arrival-time predictions.

\subsection{Limitations and Follow Ups}
In this study, the background solar wind conditions are kept fixed across the CME ensemble. This allows the influence of the prescribed CME physical parameters to be isolated, but it does not account for the variations due to the ambient solar wind. Previous studies have demonstrated that the ambient solar wind state can play an important role in determining CME evolution and arrival-time predictions \citep{case2008, Temmer2011, Lee2015, Riley2021}. Therefore, incorporating the variability of the background solar wind represents an important extension of the present work. A follow-up study is being undertaken to systematically investigate the influence of the ambient solar wind on CME evolution and to quantify its contribution to the uncertainties in CME forecasting.

\section{Conclusions}
\label{sec:Conclusions}

We investigated the influence of the physical CME properties prescribed at 0.1 AU on their subsequent heliospheric evolution and geoeffectiveness at Earth. An ensemble of 400 global MHD simulations was constructed by systematically varying the initial density, temperature, speed, and magnetic flux while retaining the same CME morphology and background solar wind. The resulting plasma, propagation and geomagnetic quantities were evaluated using correlation and variance-based sensitivity analyses to establish the relative influence of the four input parameters.

The results reveal a clear hierarchy among the CME input properties. Initial speed is the primary control on the peak radial velocity and arrival time at L1. Initial density governs the peak density, CME duration, and magnetopause compression. Magnetic flux is the principal control on the magnetic field strength at L1, while the Dst response is jointly influenced by magnetic flux and density. Temperature exhibits only weak sensitivity across the investigated outputs. Our results indicate that the primary parameters only accounts for half of the variance of their respective outputs, with other parameters taken together still contributing a non-negligible impact to total variance. Within the simulation ensemble, mean arrival-time sensitivities are 2.43~h per ($100~\mathrm{km\,s^{-1}}$) and 1.99~h per ($10^{-18}~\mathrm{kg\,m^{-3}}$) to variations in CME injection speed and density, respectively.

The ensemble also exhibits a consistent two-phase volumetric evolution, characterized by an initial period of gradual expansion followed by a systematic power-law regime. The median expansion index is $\gamma=2.38^{+0.16}_{-0.33}$. The volume-averaged magnetic, kinetic, and thermal energy densities decrease with median decay indices of 3.30, 2.67, and 2.92, respectively. The comparatively rapid decrease in magnetic energy density is broadly consistent with magnetic field dilution during CME expansion.

The sensitivity hierarchy obtained here applies to the adopted parameter ranges, fixed CME morphology and magnetic configuration, and a single structured solar wind background. In addition, the magnetopause and Dst responses are derived from empirical formulations. Future ensembles incorporating uncertainties in CME geometry, magnetic orientation, and ambient solar wind conditions will be required to determine how these factors modify the present rankings.

The results highlight the need for improved observational constraints on CME density, which remains poorly constrained despite its influence on plasma properties at L1 and the subsequent impact at Earth. Techniques such as heliospheric imaging \citep{tappin_interplanetary_2009, gibson_polarization_2026}, polarized coronagraph \citep{lamy_coronal_2020, barnes_polarimetric_2026} and radio observations \citep{xiong_interplanetary_2023,jackson_iterative_2020} can provide crucial constraints on initial CME mass and density. Improved density estimates, together with their consistent specification at the heliospheric inner boundary, are therefore essential for reducing uncertainties in predictions of CME evolution and its impact at Earth.

\section*{Acknowledgement}
JJA would like to express gratitude for the financial support received through the Prime Minister’s Research Fellowship.
PM's research was supported by the NASA Living with a Star Jack Eddy Postdoctoral Fellowship Program, administered by UCAR’s Cooperative Programs for the Advancement of Earth System Science (CPAESS) under award \#80NSSC22M0097.  PM's work was partially supported by NASA under awards No. 80NSSC23M0192, 80NSSC20K1580, 80NSSC21K1555. 
\appendix
\section{Morris sampling}
\label{app:Morris}

This appendix describes the procedure of generating Morris sampling trajectories for the input CME parameters.
A Morris sampling strategy is defined by the number of parameters ($k$) and choice of number of levels ($p$).
The procedure is as follows:

\begin{enumerate}
    \item \textbf{Define parameter ranges:}  
    Specify the physical ranges of the input parameters (density, temperature, speed, and flux).

    \item \textbf{Normalize parameters:}  
    Map the parameters to the unit interval $[0,1]$.

    \item \textbf{Discretize the parameter space:}  
    Divide each parameter into $p = 10$ levels, forming a uniform grid in the normalized space.

    \item \textbf{Generate Morris trajectories:}  
    For each trajectory:
    \begin{enumerate}
        \item Select a random starting point from the grid.
        \item Generate a sequence of $(k+1)$ points by perturbing one parameter at a time using the step size $\Delta$, where 
        \begin{equation}
            \Delta = \frac{p}{2(p-1)} \approx 0.556
        \end{equation}
    \end{enumerate}

    \item \textbf{Scale to physical space:}  
    Transform the sampled points back to their physical ranges. The flux parameter is mapped using a logarithmic transformation.

    \item \textbf{Model evaluation:}  
    Evaluate the model at each sampled point to obtain the corresponding outputs.

\end{enumerate}

\section{Sobol variance decomposition}
\label{app:Sobol}

The Sobol method is a variance-based global sensitivity analysis technique that decomposes the total variance of a computational model's output into fractions attributable to individual inputs and their joint interactions. 

For a model with $k$ independent input parameters $X = (X_1, X_2, \dots, X_k)$ producing a scalar output $Y = f(X)$, the total unconditional variance $V(Y)$ can be expanded as a sum of partial variances:

\begin{equation}
    V(Y) = \sum_{i=1}^{k} V_i + \sum_{i=1}^{k} \sum_{j=i+1}^{k} V_{ij} + \dots + V_{1,2,\dots,k}
\end{equation}

where $V_i$ represents the variance contributed by the $i$-th parameter alone, $V_{ij}$ is the variance contributed by the interaction between parameters $X_i$ and $X_j$, and higher-order terms represent multi-parameter interactions.

The first-order sensitivity index ($S_1$) represents the effect using the corresponding variable alone. It is defined as the ratio of the main effect variance to the total variance:

\begin{equation}
    S_1 = \frac{V_i}{V(Y)} = \frac{V_{X_i}(E_{X_{\sim i}}(Y | X_i))}{V(Y)}
\end{equation}

where $X_{\sim i}$ denotes the set of all input parameters except the variable of interest ($X_i$), and $E(\cdot)$ and $V(\cdot)$ denote the expected value and variance operators, respectively. The metric $S_1$ quantifies the fractional reduction in the output variance that would be achieved, on average, if the exact value of the corresponding variable could be fixed.

The total-order sensitivity index ($S_T$) represents the total variable effect. It quantifies the overall effect of a parameter, inclusive of its main effect and all higher-order interactions with other parameters. It is defined as:

\begin{equation}
S_T = 1 - \frac{V_{X_{\sim i}}(E_{X_i}(Y | X_{\sim i}))}{V(Y)}    
\end{equation}

If $S_T$ is significantly larger than $S_1$, it indicates that the corresponding variable is highly involved in complex interactions with other input parameters to drive changes in the output.

\begin{table*}[htbp]
\centering
\caption{First-order ($S_1$) and total-order ($S_T$) Sobol
sensitivity indices (\%) for the model outputs. Each entry is
reported as $S_1/S_T$.}
\label{tab:sobol}
\small
\setlength{\tabcolsep}{8pt}
\begin{tabular}{lcccc}
\hline
Output & DENSITY & SPEED & FLUX & TEMP \\
\hline
$\rho_{\max}$ & $54/76$ & $11/30$ & $3/26$ & $0/0$ \\
$v_{r,\max}$ & $36/47$ & $52/63$ & $1/1$ & $0/0$ \\
$B_{\mathrm{mag},\max}$ & $34/46$ & $6/13$ & $45/59$ & $0/0$ \\
Arrival time & $21/30$ & $40/59$ & $19/31$ & $0/0$ \\
CME duration & $35/50$ & $28/42$ & $21/26$ & $0/2$ \\
$\min(R_0/R_E)$ & $60/67$ & $30/38$ & $1/4$ & $0/0$ \\
$Dst_{\min}$ & $51/69$ & $9/21$ & $20/32$ & $1/1$ \\
$Dst_{\mathrm{area}}$ & $43/53$ & $6/16$ & $35/51$ & $0/0$ \\
\hline
\end{tabular}
\begin{flushleft}
\footnotesize
\end{flushleft}
\end{table*}

\section{Training and Validation of Surrogate Models}
\label{app:RF}
A separate Random Forest regression model was trained for each output quantity. The dataset was divided into training and testing subsets using an 80:20 split. The training subset was used to fit the surrogate, while the independent testing subset was used to evaluate its predictive performance. Each Random Forest model consisted of 1000 decision trees, with no
explicit limit on the maximum tree depth and a minimum of two samples per leaf node. For a given input vector $\mathbf{x}$, the
Random Forest prediction is obtained by averaging the predictions of the individual trees:
\begin{equation}
    \hat{f}(\mathbf{x}) =
    \frac{1}{N_{\mathrm{tree}}}
    \sum_{j=1}^{N_{\mathrm{tree}}}
    f_j(\mathbf{x}),
    \label{eq:rf_prediction}
\end{equation}
where $\hat{f}(\mathbf{x})$ denotes the surrogate prediction,
$f_j(\mathbf{x})$ is the prediction of the $j$-th decision tree, and $N_{\mathrm{tree}}=1000$ is the total number of trees in the ensemble. This ensemble-based approach allows the surrogate models to represent nonlinear relationships between the input parameters and the model outputs. Predictive performance was assessed using the coefficient of
determination ($R^2$) and the RMSE on the
held-out test subset. These metrics are defined as
\begin{equation}
    R^2 = 1 - \frac{\sum_{i=1}^{n_{\mathrm{test}}}
    (y_i-\hat{y}_i)^2}{\sum_{i=1}^{n_{\mathrm{test}}}
    (y_i-\bar{y})^2},
    \label{eq:surrogate_r2}
\end{equation}
and
\begin{equation}
    \mathrm{RMSE} =\sqrt{
    \frac{1}{n_{\mathrm{test}}} \sum_{i=1}^{n_{\mathrm{test}}} (y_i-\hat{y}_i)^2},
    \label{eq:surrogate_rmse}
\end{equation}
where $y_i$ and $\hat{y}_i$ are the simulated and predicted output values, respectively, $\bar{y}$ is the mean of the observed test outputs, and $n_{\mathrm{test}}$ is the number of test samples.
\begin{table}[h]
\centering
\caption{Performance metrics of the Random Forest surrogate models used for Sobol sensitivity analysis. The table lists the test-set $R^2$, RMSE, and the mean and CV $R^2$.}
\label{tab:surrogate_performance}
\begin{tabular}{lccc}
\hline
\textbf{Output} & \textbf{Test $R^2$} & \textbf{RMSE} & \textbf{CV $R^2$} \\
\hline
$\rho_{\max}$ & 0.917 & 5.505 & $0.943 \pm 0.020$ \\
$v_{r,\max}$ & 0.984 & 24.259 & $0.989 \pm 0.004$ \\
$B_{\mathrm{mag},\max}$ & 0.981 & 1.568 & $0.976 \pm 0.010$ \\
CME time of arrival & 0.965 & 2.545 & $0.945 \pm 0.016$ \\
CME duration & 0.975 & 2.815 & $0.975 \pm 0.013$ \\
$r_{0,\min}$ & 0.977 & 0.165 & $0.974 \pm 0.007$ \\
Dst$_{\min}$ & 0.942 & 11.820 & $0.931 \pm 0.026$ \\
Dst area & 0.934 & 361.068 & $0.932 \pm 0.020$ \\
\hline
\end{tabular}
\end{table}
A higher $R^2$ and a lower RMSE indicate better predictive performance. In addition to the hold-out evaluation, five-fold cross-validation (CV $R^2$) was performed. The dataset was divided into five approximately equal subsets, with each subset used once for validation while the remaining subsets were used for training. The mean and standard deviation of the resulting $R^2$ scores were recorded to assess predictive performance across different data partitions. The test-set metrics and cross-validation scores for each output parameter is shown in Table~\ref{tab:surrogate_performance}.

\bibliography{bib.bib}

@article{Morris1991,
  title={Factorial sampling plans for preliminary computational experiments},
  author={Morris, Max D.},
  journal={Technometrics},
  volume={33},
  number={2},
  pages={161--174},
  year={1991},
  publisher={Taylor \& Francis}
}

@ARTICLE{Issan2023,
       author = {{Issan}, Opal and {Riley}, Pete and {Camporeale}, Enrico and {Kramer}, Boris},
        title = "{Bayesian Inference and Global Sensitivity Analysis for Ambient Solar Wind Prediction}",
      journal = {Space Weather},
         year = 2023,
        month = sep,
       volume = {21},
       number = {9},
          eid = {e2023SW003555},
        pages = {e2023SW003555},
          doi = {10.1029/2023SW003555},
archivePrefix = {arXiv},
       eprint = {2305.08009},
 primaryClass = {physics.space-ph},
       adsurl = {https://ui.adsabs.harvard.edu/abs/2023SpWea..2103555I}
}

@ARTICLE{OBrien2000,
       author = {{O'Brien}, T.~P. and {McPherron}, R.~L.},
        title = "{Forecasting the ring current index Dst in real time}",
      journal = {Journal of Atmospheric and Solar-Terrestrial Physics},
         year = 2000,
        month = sep,
       volume = {62},
       number = {14},
        pages = {1295-1299},
          doi = {10.1016/S1364-6826(00)00072-9},
       adsurl = {https://ui.adsabs.harvard.edu/abs/2000JASTP..62.1295O}
}

@ARTICLE{Shue1998,
       author = {{Shue}, J.-H. and {Song}, P. and {Russell}, C.~T. and {Steinberg}, J.~T. and {Chao}, J.~K. and {Zastenker}, G. and {Vaisberg}, O.~L. and {Kokubun}, S. and {Singer}, H.~J. and {Detman}, T.~R. and {Kawano}, H.},
        title = "{Magnetopause location under extreme solar wind conditions}",
      journal = {\jgr},
         year = 1998,
        month = aug,
       volume = {103},
       number = {A8},
        pages = {17691-17700},
          doi = {10.1029/98JA01103},
       adsurl = {https://ui.adsabs.harvard.edu/abs/1998JGR...10317691S}
}

@Article{Mayank2023,
  author    = {Mayank, Prateek and Vaidya, Bhargav and Mishra, Wageesh and Chakrabarty, D.},
  journal   = {The Astrophysical Journal Supplement Series},
  title     = {SWASTi-CME: A Physics-based Model to Study Coronal Mass Ejection Evolution and Its Interaction with Solar Wind},
  year      = {2023},
  issn      = {1538-4365},
  month     = dec,
  number    = {1},
  pages     = {10},
  volume    = {270},
  doi       = {10.3847/1538-4365/ad08c7},
  publisher = {American Astronomical Society},
}

@Article{Mayank2022,
  author    = {Mayank, Prateek and Vaidya, Bhargav and Chakrabarty, D.},
  journal   = {The Astrophysical Journal Supplement Series},
  title     = {SWASTi-SW: Space Weather Adaptive Simulation Framework for Solar Wind and Its Relevance to the Aditya-L1 Mission},
  year      = {2022},
  issn      = {1538-4365},
  month     = sep,
  number    = {1},
  pages     = {23},
  volume    = {262},
  doi       = {10.3847/1538-4365/ac8551},
  publisher = {American Astronomical Society},
}

@ARTICLE{Mayank2025,
  author  = {Mayank, Prateek and J. Athalathil, Jithu and Nandy, Sirsha and Vaidya, Bhargav and Navanit, A. V. and Paul, Arghyadeep},
  title   = "{SWASTi: A physics-based modelling toolkit for space weather}",
  journal = {Journal of Astrophysics and Astronomy},
  year    = {2025},
  month   = {oct},
  volume  = {46},
  number  = {2},
  eid     = {80},
  pages   = {80},
  doi     = {10.1007/s12036-025-10107-2},
  adsurl  = {https://ui.adsabs.harvard.edu/abs/2025JApA...46...80M}
}

@InProceedings{Arge2003,
  author    = {Arge, Charles N.},
  booktitle = {AIP Conference Proceedings},
  title     = {Improved Method for Specifying Solar Wind Speed Near the Sun},
  year      = {2003},
  pages     = {190--193},
  publisher = {AIP},
  volume    = {679},
  doi       = {10.1063/1.1618574},
  issn      = {0094-243X},
}

@Article{Stansby2020,
  author    = {Stansby, David and Yeates, Anthony and Badman, Samuel},
  journal   = {Journal of Open Source Software},
  title     = {pfsspy: A Python package for potential field source surface modelling},
  year      = {2020},
  issn      = {2475-9066},
  month     = oct,
  number    = {54},
  pages     = {2732},
  volume    = {5},
  doi       = {10.21105/joss.02732},
  publisher = {The Open Journal},
}

@ARTICLE{Ballegooijen2000,
       author = {{van Ballegooijen}, A.~A. and {Priest}, E.~R. and {Mackay}, D.~H.},
        title = "{Mean Field Model for the Formation of Filament Channels on the Sun}",
      journal = {\apj},
         year = 2000,
        month = aug,
       volume = {539},
       number = {2},
        pages = {983-994},
          doi = {10.1086/309265},
       adsurl = {https://ui.adsabs.harvard.edu/abs/2000ApJ...539..983V}
}

@Article{Isavnin2016,
  author    = {Isavnin, A.},
  journal   = {The Astrophysical Journal},
  title     = {FRiED: A NOVEL THREE-DIMENSIONAL MODEL OF CORONAL MASS EJECTIONS},
  year      = {2016},
  issn      = {1538-4357},
  month     = dec,
  number    = {2},
  pages     = {267},
  volume    = {833},
  doi       = {10.3847/1538-4357/833/2/267},
  publisher = {American Astronomical Society},
}

@ARTICLE{Falkenberg2010,
       author = {{Falkenberg}, T.~V. and {Vr{\v{s}}nak}, B. and {Taktakishvili}, A. and {Odstrcil}, D. and {MacNeice}, P. and {Hesse}, M.},
        title = "{Investigations of the sensitivity of a coronal mass ejection model (ENLIL) to solar input parameters}",
      journal = {Space Weather},
         year = 2010,
        month = jun,
       volume = {8},
       number = {6},
          eid = {S06004},
        pages = {S06004},
          doi = {10.1029/2009SW000555},
       adsurl = {https://ui.adsabs.harvard.edu/abs/2010SpWea...8.6004F}
}

@ARTICLE{Koehn2022,
       author = {{Koehn}, G.~J. and {Desai}, R.~T. and {Davies}, E.~E. and {Forsyth}, R.~J. and {Eastwood}, J.~P. and {Poedts}, S.},
        title = "{Successive Interacting Coronal Mass Ejections: How to Create a Perfect Storm}",
      journal = {\apj},
         year = 2022,
        month = dec,
       volume = {941},
       number = {2},
          eid = {139},
        pages = {139},
          doi = {10.3847/1538-4357/aca28c},
archivePrefix = {arXiv},
       eprint = {2211.05899},
 primaryClass = {astro-ph.SR},
       adsurl = {https://ui.adsabs.harvard.edu/abs/2022ApJ...941..139K}
}

@ARTICLE{Lee2015,
       author = {{Lee}, C.~O. and {Arge}, C.~N. and {Odstrcil}, D. and {Millward}, G. and {Pizzo}, V. and {Lugaz}, N.},
        title = "{Ensemble Modeling of Successive Halo CMEs: A Case Study}",
      journal = {\solphys},
         year = 2015,
        month = apr,
       volume = {290},
       number = {4},
        pages = {1207-1229},
          doi = {10.1007/s11207-015-0667-2},
       adsurl = {https://ui.adsabs.harvard.edu/abs/2015SoPh..290.1207L}
}

@ARTICLE{Riley2021,
       author = {{Riley}, Pete and {Ben-Nun}, Michal},
        title = "{On the Sources and Sizes of Uncertainty in Predicting the Arrival Time of Interplanetary Coronal Mass Ejections Using Global MHD Models}",
      journal = {Space Weather},
         year = 2021,
        month = jun,
       volume = {19},
       number = {6},
          eid = {e02775},
        pages = {e02775},
          doi = {10.1029/2021SW00277510.1002/essoar.10506742.1},
       adsurl = {https://ui.adsabs.harvard.edu/abs/2021SpWea..1902775R}
}

@ARTICLE{Taktakishvili2010,
       author = {{Taktakishvili}, A. and {MacNeice}, P. and {Odstrcil}, D.},
        title = "{Model uncertainties in predictions of arrival of coronal mass ejections at Earth orbit}",
      journal = {Space Weather},
         year = 2010,
        month = jun,
       volume = {8},
       number = {6},
          eid = {S06007},
        pages = {S06007},
          doi = {10.1029/2009SW000543},
       adsurl = {https://ui.adsabs.harvard.edu/abs/2010SpWea...8.6007T}
}

@ARTICLE{Wang2026,
       author = {{Wang}, Ying and {Chen}, Yu and {Zhuang}, Bin and {Xu}, Yan and {Gou}, Tingyu and {liu}, Nian and {Wang}, Haimin},
        title = "{Study of the Orientation of Erupting Magnetic Flux Ropes in Coronal Mass Ejections from the Solar Corona to 1 au}",
      journal = {\apj},
         year = 2026,
        month = feb,
       volume = {997},
       number = {2},
          eid = {359},
        pages = {359},
          doi = {10.3847/1538-4357/ae2c7a},
       adsurl = {https://ui.adsabs.harvard.edu/abs/2026ApJ...997..359W}
}

@ARTICLE{Gosling1990,
       author = {{Gosling}, J.~T. and {Bame}, S.~J. and {McComas}, D.~J. and {Phillips}, J.~L.},
        title = "{Coronal mass ejections and large geomagnetic storms}",
      journal = {\grl},
         year = 1990,
        month = jun,
       volume = {17},
       number = {7},
        pages = {901-904},
          doi = {10.1029/GL017i007p00901},
       adsurl = {https://ui.adsabs.harvard.edu/abs/1990GeoRL..17..901G}
}

@ARTICLE{Odstrcil2004,
       author = {{Odstrcil}, D. and {Riley}, P. and {Zhao}, X.~P.},
        title = "{Numerical simulation of the 12 May 1997 interplanetary CME event}",
      journal = {Journal of Geophysical Research (Space Physics)},
         year = 2004,
        month = feb,
       volume = {109},
       number = {A2},
          eid = {A02116},
        pages = {A02116},
          doi = {10.1029/2003JA010135},
       adsurl = {https://ui.adsabs.harvard.edu/abs/2004JGRA..109.2116O}
}

@ARTICLE{Shiota2016,
       author = {{Shiota}, D. and {Kataoka}, R.},
        title = "{Magnetohydrodynamic simulation of interplanetary propagation of multiple coronal mass ejections with internal magnetic flux rope (SUSANOO-CME)}",
      journal = {Space Weather},
         year = 2016,
        month = feb,
       volume = {14},
       number = {2},
        pages = {56-75},
          doi = {10.1002/2015SW001308},
       adsurl = {https://ui.adsabs.harvard.edu/abs/2016SpWea..14...56S}
}

@ARTICLE{Pomoell2018,
       author = {{Pomoell}, Jens and {Poedts}, S.},
        title = "{EUHFORIA: European heliospheric forecasting information asset}",
      journal = {Journal of Space Weather and Space Climate},
         year = 2018,
        month = jun,
       volume = {8},
          eid = {A35},
        pages = {A35},
          doi = {10.1051/swsc/2018020},
       adsurl = {https://ui.adsabs.harvard.edu/abs/2018JSWSC...8A..35P}
}

@INPROCEEDINGS{Odstrcil2003,
       author = {{Odstrcil}, D. and {Riley}, P. and {Linker}, J.~A. and {Lionello}, R. and {Mikic}, Z. and {Pizzo}, V.~J.},
        title = "{3-D simulations of ICMEs by coupled coronal and heliospheric models}",
    booktitle = {Solar Variability as an Input to the Earth's Environment},
         year = 2003,
       editor = {{Wilson}, Andrew},
       series = {ESA Special Publication},
       volume = {535},
        month = sep,
        pages = {541-546},
       adsurl = {https://ui.adsabs.harvard.edu/abs/2003ESASP.535..541O}
}

@article{Barnard2020,
author = {Barnard, L. and Owens, M. J. and Scott, C. J. and de Koning, C. A.},
title = {Ensemble CME Modeling Constrained by Heliospheric Imager Observations},
journal = {AGU Advances},
volume = {1},
number = {3},
pages = {e2020AV000214},
doi = {https://doi.org/10.1029/2020AV000214},
url = {https://agupubs.onlinelibrary.wiley.com/doi/abs/10.1029/2020AV000214},
eprint = {https://agupubs.onlinelibrary.wiley.com/doi/pdf/10.1029/2020AV000214},
note = {e2020AV000214 10.1029/2020AV000214},
year = {2020}
}

@ARTICLE{Temmer2011,
       author = {{Temmer}, Manuela and {Rollett}, Tanja and {M{\"o}stl}, Christian and {Veronig}, Astrid M. and {Vr{\v{s}}nak}, Bojan and {Odstr{\v{c}}il}, Dusan},
        title = "{Influence of the Ambient Solar Wind Flow on the Propagation Behavior of Interplanetary Coronal Mass Ejections}",
      journal = {\apj},
         year = 2011,
        month = dec,
       volume = {743},
       number = {2},
          eid = {101},
        pages = {101},
          doi = {10.1088/0004-637X/743/2/101},
archivePrefix = {arXiv},
       eprint = {1110.0827},
 primaryClass = {astro-ph.SR},
       adsurl = {https://ui.adsabs.harvard.edu/abs/2011ApJ...743..101T}
}

@article{RichardsonCane2010,
  author  = {Richardson, I. G. and Cane, H. V.},
  title   = {Near-Earth Interplanetary Coronal Mass Ejections During
             Solar Cycle 23 (1996--2009): Catalog and Summary of Properties},
  journal = {Solar Physics},
  volume  = {264},
  pages   = {189--237},
  year    = {2010},
  doi     = {10.1007/s11207-010-9568-6}
}

@article{RichardsonCane2011,
  author  = {Richardson, I. G. and Cane, H. V.},
  title   = {Geoeffectiveness ({Dst} and {Kp}) of Interplanetary
             Coronal Mass Ejections During 1995--2009 and
             Implications for Storm Forecasting},
  journal = {Space Weather},
  volume  = {9},
  pages   = {S07005},
  year    = {2011},
  doi     = {10.1029/2011SW000670}
}

@ARTICLE{Case2008,
       author = {{Case}, A.~W. and {Spence}, H.~E. and {Owens}, M.~J. and {Riley}, P. and {Odstrcil}, D.},
        title = "{Ambient solar wind's effect on ICME transit times}",
      journal = {\grl},
         year = 2008,
        month = aug,
       volume = {35},
       number = {15},
          eid = {L15105},
        pages = {L15105},
          doi = {10.1029/2008GL034493},
       adsurl = {https://ui.adsabs.harvard.edu/abs/2008GeoRL..3515105C}
}

@article{maharana_updates_2023,
	title = {Updates on the {FRi3D} {CME} model in {EUHFORIA}},
	url = {http://arxiv.org/abs/2310.11402},
	doi = {10.48550/arXiv.2310.11402},
	urldate = {2026-03-23},
	publisher = {arXiv},
	author = {Maharana, Anwesha and Plets, Karel and Isavnin, Alexey and Poedts, Stefaan},
	month = oct,
	year = {2023},
	note = {arXiv:2310.11402 [astro-ph]},
}

@ARTICLE{Altschuler1969,
       author = {{Altschuler}, Martin D. and {Newkirk}, Jr., Gordon},
        title = "{Magnetic Fields and the Structure of the Solar Corona. I: Methods of Calculating Coronal Fields}",
      journal = {\solphys},
         year = 1969,
        month = sep,
       volume = {9},
       number = {1},
        pages = {131-149},
          doi = {10.1007/BF00145734},
       adsurl = {https://ui.adsabs.harvard.edu/abs/1969SoPh....9..131A}
}

@ARTICLE{Toth2012,
  author  = {{T{\'o}th}, G{\'a}bor and {van der Holst}, Bart and {Sokolov}, Igor V. and
             {De Zeeuw}, Darren L. and {Gombosi}, Tamas I. and {Fang}, Fang and
             {Manchester}, Ward B. and {Meng}, Xing and {Najib}, Dalal and
             {Powell}, Kenneth G. and {Stout}, Quentin F. and {Glocer}, Alex and
             {Ma}, Ying-Juan and {Opher}, Merav},
  title   = "{Adaptive numerical algorithms in space weather modeling}",
  journal = {Journal of Computational Physics},
  year    = {2012},
  volume  = {231},
  number  = {3},
  pages   = {870--903},
  doi     = {10.1016/j.jcp.2011.02.006}
}

@ARTICLE{Toth2005,
       author = {{T{\'o}th}, G{\'a}Bor and {Sokolov}, Igor V. and {Gombosi}, Tamas I. and {Chesney}, David R. and {Clauer}, C. Robert and {de Zeeuw}, Darren L. and {Hansen}, Kenneth C. and {Kane}, Kevin J. and {Manchester}, Ward B. and {Oehmke}, Robert C. and {Powell}, Kenneth G. and {Ridley}, Aaron J. and {Roussev}, Ilia I. and {Stout}, Quentin F. and {Volberg}, Ovsei and {Wolf}, Richard A. and {Sazykin}, Stanislav and {Chan}, Anthony and {Yu}, Bin and {K{\'o}ta}, J{\'o}zsef},
        title = "{Space Weather Modeling Framework: A new tool for the space science community}",
      journal = {Journal of Geophysical Research (Space Physics)},
         year = 2005,
        month = dec,
       volume = {110},
       number = {A12},
          eid = {A12226},
        pages = {A12226},
          doi = {10.1029/2005JA011126},
       adsurl = {https://ui.adsabs.harvard.edu/abs/2005JGRA..11012226T}
}

@ARTICLE{vanderHolst2014,
  author  = {{van der Holst}, B. and {Sokolov}, I.~V. and {Meng}, X. and {Jin}, M. and
             {Manchester}, W.~B., IV and {T{\'o}th}, G. and {Gombosi}, T.~I.},
  title   = "{Alfv{\'e}n Wave Solar Model (AWSoM): Coronal Heating}",
  journal = {The Astrophysical Journal},
  year    = {2014},
  volume  = {782},
  number  = {2},
  eid     = {81},
  pages   = {81},
  doi     = {10.1088/0004-637X/782/2/81}
}

@ARTICLE{Mikic1999,
  author  = {{Miki{\'c}}, Z. and {Linker}, J.~A. and {Schnack}, D.~D. and
             {Lionello}, R. and {Tarditi}, A.},
  title   = "{Magnetohydrodynamic modeling of the global solar corona}",
  journal = {Physics of Plasmas},
  year    = {1999},
  volume  = {6},
  number  = {5},
  pages   = {2217--2224},
  doi     = {10.1063/1.873474}
}

@ARTICLE{Riley2001,
  author  = {{Riley}, Pete and {Linker}, J.~A. and {Miki{\'c}}, Z.},
  title   = "{An empirically-driven global MHD model of the solar corona and inner heliosphere}",
  journal = {Journal of Geophysical Research},
  year    = {2001},
  volume  = {106},
  number  = {A8},
  pages   = {15889--15901},
  doi     = {10.1029/2000JA000121}
}

@ARTICLE{Merkin2016,
  author  = {{Merkin}, V.~G. and {Lyon}, J.~G. and {Lario}, D. and
             {Arge}, C.~N. and {Henney}, C.~J.},
  title   = "{Time-dependent magnetohydrodynamic simulations of the inner heliosphere}",
  journal = {Journal of Geophysical Research: Space Physics},
  year    = {2016},
  volume  = {121},
  number  = {4},
  pages   = {2866--2890},
  doi     = {10.1002/2015JA022200}
}

@ARTICLE{Provornikova2024,
  author  = {{Provornikova}, E. and {Merkin}, V.~G. and {Vourlidas}, A. and
             {Malanushenko}, A. and {Gibson}, S.~E. and {Winter}, E. and {Arge}, C.~N.},
  title   = "{MHD Modeling of a Geoeffective Interplanetary Coronal Mass Ejection with the Magnetic Topology Informed by In Situ Observations}",
  journal = {The Astrophysical Journal},
  year    = {2024},
  volume  = {977},
  number  = {1},
  eid     = {106},
  pages   = {106},
  doi     = {10.3847/1538-4357/ad83b1}
}

@ARTICLE{Verbeke2022,
  author  = {{Verbeke}, C. and {Baratashvili}, T. and {Poedts}, S.},
  title   = "{Icarus: A new inner heliospheric model with a flexible grid}",
  journal = {Astronomy \& Astrophysics},
  year    = {2022},
  volume  = {662},
  eid     = {A50},
  pages   = {A50},
  doi     = {10.1051/0004-6361/202141981}
}

@article{Linker2024,
  author  = {Linker, Jon A. and Torok, Tibor and Downs, Cooper and Caplan, Ronald M. and Titov, Viacheslav and Reyes, Andres and Lionello, Roberto and Riley, Pete},
  title   = {{CORHEL-CME}: An Interactive Tool For Modeling Solar Eruptions},
  journal = {Journal of Physics: Conference Series},
  year    = {2024},
  volume  = {2742},
  number  = {1},
  pages   = {012012},
  doi     = {10.1088/1742-6596/2742/1/012012}
}

@ARTICLE{Verbeke2023,
  author  = {{Verbeke}, C. and {Mays}, M.~L. and {Kay}, C. and {Riley}, P. and
             {Palmerio}, E. and {Dumbovi{\'c}}, M. and {Mierla}, M. and
             {Scolini}, C. and {Temmer}, M. and {Paouris}, E. and
             {Jensen}, E.~A. and {Bisi}, M.~M. and {Hinterreiter}, J.},
  title   = "{Quantifying errors in 3D CME parameters derived from synthetic data using white-light reconstruction techniques}",
  journal = {Advances in Space Research},
  year    = {2023},
  volume  = {72},
  number  = {12},
  pages   = {5243--5262},
  doi     = {10.1016/j.asr.2022.08.056}
}

@ARTICLE{Mays2015,
  author  = {{Mays}, M.~L. and {Taktakishvili}, A. and {Pulkkinen}, A. and
             {MacNeice}, P.~J. and {Rast{\"a}tter}, L. and {Odstrcil}, D. and
             {Jian}, L.~K. and {Richardson}, I.~G. and {LaSota}, J.~A. and
             {Zheng}, Y. and {Kuznetsova}, M.~M.},
  title   = "{Ensemble Modeling of CMEs Using the WSA-ENLIL+Cone Model}",
  journal = {Solar Physics},
  year    = {2015},
  volume  = {290},
  number  = {6},
  pages   = {1775--1814},
  doi     = {10.1007/s11207-015-0692-1}
}

@ARTICLE{Scolini2018,
  author  = {{Scolini}, C. and {Verbeke}, C. and {Poedts}, S. and {Chan{\'e}}, E. and
             {Pomoell}, J. and {Zuccarello}, F.~P.},
  title   = "{Effect of the Initial Shape of Coronal Mass Ejections on 3-D MHD Simulations and Geoeffectiveness Predictions}",
  journal = {Space Weather},
  year    = {2018},
  volume  = {16},
  number  = {6},
  pages   = {754--771},
  doi     = {10.1029/2018SW001806}
}

@ARTICLE{Dumbovic2018,
  author  = {{Dumbovi{\'c}}, Mateja and {{\v C}alogovi{\'c}}, Ja{\v s}a and
             {Vr{\v s}nak}, Bojan and {Temmer}, Manuela and {Mays}, M. Leila and
             {Veronig}, Astrid and {Piantschitsch}, Isabell},
  title   = "{The Drag-based Ensemble Model (DBEM) for Coronal Mass Ejection Propagation}",
  journal = {The Astrophysical Journal},
  year    = {2018},
  volume  = {854},
  number  = {2},
  eid     = {180},
  pages   = {180},
  doi     = {10.3847/1538-4357/aaaa66}
}

@ARTICLE{Mierla2010,
  author  = {{Mierla}, M. and {Inhester}, B. and {Antunes}, A. and {Boursier}, Y. and
             {Byrne}, J.~P. and {Colaninno}, R. and {Davila}, J. and {de Koning}, C.~A. and
             {Gallagher}, P.~T. and {Gissot}, S. and {Howard}, R.~A. and {Howard}, T.~A. and
             {Kramar}, M. and {Lamy}, P. and {Liewer}, P.~C. and {Maloney}, S. and
             {Marqu{\'e}}, C. and {McAteer}, R.~T.~J. and {Moran}, T. and {Rodriguez}, L. and
             {Srivastava}, N. and {St. Cyr}, O.~C. and {Stenborg}, G. and {Temmer}, M. and
             {Thernisien}, A. and {Vourlidas}, A. and {West}, M.~J. and {Wood}, B.~E. and
             {Zhukov}, A.~N.},
  title   = "{On the 3-D reconstruction of Coronal Mass Ejections using coronagraph data}",
  journal = {Annales Geophysicae},
  year    = {2010},
  volume  = {28},
  number  = {1},
  pages   = {203--215},
  doi     = {10.5194/angeo-28-203-2010}
}

@article{xiong_interplanetary_2023,
	title = {Interplanetary scintillation observation and space weather modelling},
	volume = {10},
	issn = {2296-987X},
	url = {https://www.frontiersin.org/journals/astronomy-and-space-sciences/articles/10.3389/fspas.2023.1159166/full},
	doi = {10.3389/fspas.2023.1159166},
	language = {English},
	urldate = {2026-08-19},
	journal = {Frontiers in Astronomy and Space Sciences},
	publisher = {Frontiers},
	author = {Xiong, Ming and Feng, Xueshang and Li, Bo and He, Jiansen and Wang, Wei and Gao, Yanchen and Zhang, Man and Yang, Liping and Huang, Zhenghua and Cheng, Jun and Su, Cang and Yan, Yihua and Ying, Kairan},
	month = jul,
	year = {2023},
}

@article{jackson_iterative_2020,
	title = {Iterative {Tomography}: {A} {Key} to {Providing} {Time}-{Dependent} 3-{D} {Reconstructions} of the {Inner} {Heliosphere} and the {Unification} of {Space} {Weather} {Forecasting} {Techniques}},
	volume = {7},
	issn = {2296-987X},
	shorttitle = {Iterative {Tomography}},
	url = {https://www.frontiersin.org/journals/astronomy-and-space-sciences/articles/10.3389/fspas.2020.568429/full},
	doi = {10.3389/fspas.2020.568429},
	language = {English},
	urldate = {2026-09-02},
	journal = {Frontiers in Astronomy and Space Sciences},
	publisher = {Frontiers},
	author = {Jackson, Bernard V. and Buffington, Andrew and Cota, Lucas and Odstrcil, Dusan and Bisi, Mario M. and Fallows, Richard and Tokumaru, Munetoshi},
	month = nov,
	year = {2020},
}

@article{lamy_coronal_2020,
	title = {Coronal {Photopolarimetry} with the {LASCO}-{C2} {Coronagraph} over 24 {Years} [1996 – 2019]: {Application} to the {K}/{F} {Separation} and to the {Determination} of the {Electron} {Density}},
	volume = {295},
	issn = {0038-0938, 1573-093X},
	shorttitle = {Coronal {Photopolarimetry} with the {LASCO}-{C2} {Coronagraph} over 24 {Years} [1996 – 2019]},
	url = {https://link.springer.com/10.1007/s11207-020-01650-y},
	doi = {10.1007/s11207-020-01650-y},
	language = {en},
	number = {7},
	urldate = {2025-09-16},
	journal = {Solar Physics},
	author = {Lamy, Philippe and Llebaria, Antoine and Boclet, Brice and Gilardy, Hugo and Burtin, Michael and Floyd, Olivier},
	month = jul,
	year = {2020},
	pages = {89},
}

@article{barnes_polarimetric_2026,
	title = {Polarimetric tomography applied to synthetic multi-spacecraft white-light images: observing coronal mass ejections in {3D}},
	volume = {548},
	issn = {0035-8711},
	shorttitle = {Polarimetric tomography applied to synthetic multi-spacecraft white-light images},
	url = {https://doi.org/10.1093/mnras/stag818},
	doi = {10.1093/mnras/stag818},
	number = {4},
	urldate = {2026-07-31},
	journal = {Monthly Notices of the Royal Astronomical Society},
	author = {Barnes, David and Palmerio, Erika and Amerstorfer, Tanja and Asvestari, Eleanna and Barnard, Luke and Bauer, Maike and Čalogović, Jaša and Cappello, Greta M and Hess, Phillip and Kay, Christina},
	month = jun,
	year = {2026},
	pages = {stag818},
}

@article{tappin_interplanetary_2009,
	title = {Interplanetary {Coronal} {Mass} {Ejections} {Observed} in the {Heliosphere}: 2. {Model} and {Data} {Comparison}},
	volume = {147},
	issn = {1572-9672},
	shorttitle = {Interplanetary {Coronal} {Mass} {Ejections} {Observed} in the {Heliosphere}},
	url = {https://doi.org/10.1007/s11214-009-9550-5},
	doi = {10.1007/s11214-009-9550-5},
	language = {en},
	number = {1},
	urldate = {2026-01-30},
	journal = {Space Science Reviews},
	author = {Tappin, S. James and Howard, Timothy A.},
	month = oct,
	year = {2009},
	pages = {55--87},
}

@article{gibson_polarization_2026,
	title = {Polarization {Diagnostics} {Applied} to {Coronal} {Mass} {Ejections} and the {Background} {Solar} {Wind}},
	volume = {301},
	issn = {1573-093X},
	url = {https://doi.org/10.1007/s11207-026-02620-6},
	doi = {10.1007/s11207-026-02620-6},
	language = {en},
	number = {2},
	urldate = {2026-08-02},
	journal = {Solar Physics},
	author = {Gibson, Sarah E. and DeForest, Craig E. and de Koning, Curt A. and Cranmer, Steven R. and Fan, Yuhong and Morgan, Huw and Provornikova, Elena and Malanushenko, Anna and Webb, David},
	month = feb,
	year = {2026},
	pages = {33},
}

@ARTICLE{Millas2020,
       author = {{Millas}, Dimitrios and {Innocenti}, Maria Elena and {Laperre}, Brecht and {Raeder}, Joachim and {Poedts}, Stefaan and {Lapenta}, Giovanni},
        title = "{Domain of Influence analysis: implications for Data Assimilation in space weather forecasting}",
      journal = {Frontiers in Astronomy and Space Sciences},
         year = 2020,
        month = oct,
       volume = {7},
          eid = {73},
        pages = {73},
          doi = {10.3389/fspas.2020.571286},
archivePrefix = {arXiv},
       eprint = {2009.04211},
 primaryClass = {physics.space-ph},
       adsurl = {https://ui.adsabs.harvard.edu/abs/2020FrASS...7...73M}
}

@ARTICLE{Moestl2026,
       author = {{M{\"o}stl}, Christian and {Davies}, Emma E. and {Weiler}, Eva and {R{\"u}disser}, Hannah T. and {Amerstorfer}, Ute V. and {Weiss}, Andreas J. and {Reiss}, Martin A. and {Majumdar}, Satabdwa and {Horbury}, Timothy S. and {Bale}, Stuart D. and et al.},
        title = "{On the Magnetic Field Evolution of Interplanetary Coronal Mass Ejections from 0.07 to 5.4 au}",
      journal = {\apj},
         year = 2026,
        month = apr,
       volume = {1001},
       number = {1},
          eid = {70},
        pages = {70},
          doi = {10.3847/1538-4357/ae50fe},
archivePrefix = {arXiv},
       eprint = {2512.04730},
 primaryClass = {astro-ph.SR},
       adsurl = {https://ui.adsabs.harvard.edu/abs/2026ApJ..1001...70M}
}
\bibliographystyle{aasjournal}

\end{document}